\documentclass[twocolumn]{aastex631}
\newcommand\ha{$\mathrm{H}\alpha$\xspace}
\newcommand\teff{$T_{\mathrm{eff}}$\xspace}
\newcommand{\beatlas}{\textsc{BeAtlas}\xspace}

\newcommand{\mycomment}[1]{}
\DeclareUnicodeCharacter{2212}{-}
\usepackage{amssymb}
\usepackage[normalem]{ulem}
\usepackage{xcolor}
\usepackage{xspace}
\usepackage{chngpage}
\usepackage{tabularx}
\usepackage{ragged2e}
\shorttitle{AO-assisted $\mli{BVRI}$+H$\alpha$ imaging of NGC\,330 with SAMI/SOAR}
\shortauthors{Navarete et al.}

\begin{document}

\title{On the Origin of Fast Rotating Stars. \\ I. Photometric calibration and results of AO-assisted $\mli{BVRI}$+H$\alpha$ imaging of NGC\,330 with SAMI/SOAR
}

\correspondingauthor{Felipe Navarete}
\email{felipe.navarete@noirlab.com}

\author[0000-0002-0284-0578]{Felipe Navarete}
\affiliation{SOAR Telescope/NSF's NOIRLab, Avda Juan Cisternas 1500, 1700000, La Serena, Chile}

\author[0000-0002-4808-7796]{Pedro Ticiani dos Santos}
\affiliation{Universidade de S\~ao Paulo, Instituto de Astronomia, Geof\'isica e Ci\^encias Atmosf\'ericas, \\ Rua do Mat\~ao 1226, Cidade Universit\'aria S\~ao Paulo-SP, 05508-090, Brazil}

\author[0000-0002-9369-574X]{Alex Cavaliéri Carciofi}
\affiliation{Universidade de S\~ao Paulo, Instituto de Astronomia, Geof\'isica e Ci\^encias Atmosf\'ericas, \\ Rua do Mat\~ao 1226, Cidade Universit\'aria S\~ao Paulo-SP, 05508-090, Brazil}

\author[0009-0007-1625-8937]{André Luiz Figueiredo}
\affiliation{Universidade de S\~ao Paulo, Instituto de Astronomia, Geof\'isica e Ci\^encias Atmosf\'ericas, \\ Rua do Mat\~ao 1226, Cidade Universit\'aria S\~ao Paulo-SP, 05508-090, Brazil}

\begin{abstract}
{\ha} emission is a clear indicator of circumstellar activity in Be stars, historically employed to assess the classical Be star (CBe) population in young open clusters (YOCs). The YOC NGC\,330 in the Small Magellanic Cloud exhibits a large known fraction of {CBe} stars and was selected for a pilot study to establish a comprehensive methodology for identifying {\ha} emitters in the Magellanic Clouds, encompassing the entire B-type spectral range. Using the SOAR Adaptative Module Imager (SAMI), we investigated the stellar population of NGC\,330 using multi-band $\mli{BVRI}$+{\ha} imaging. We identified {\ha} emitters within the entire V-band range covered by SAMI/SOAR observations ($V\lesssim22$), comprising the complete B-type stellar population and offering a unique opportunity to explore the Be phenomenon across all spectral sub-classes. The stellar radial distribution shows a clear bimodal pattern between the most massive (B5 or earlier) and the lower-mass main-sequence objects (later than B6) within the cluster. The former is concentrated towards the cluster center (showing a dispersion of $\sigma=4.26\pm0.20$~pc), whereas the latter extends across larger radii ($\sigma=10.83\pm0.65$~pc), indicating mass stratification within NGC\,330.
The total fraction of emitters is $4.4\pm0.5\%$, notably smaller than previous estimates from flux- or seeing-limited observations. However, a higher fraction of {\ha} emitters is observed among higher-mass stars ($32.8\pm3.4\%$) than within lower-mass ($4.4\pm0.9\%$). Consequently, the putative CBe population exhibits distinct dynamical characteristics compared to the bulk of the stellar population in NGC\,330.
These findings highlight the significance of the current observations in providing a complete picture of the CBe population in NGC\,330.
\end{abstract}


\section{Introduction}
\label{sec_intro}

Massive stars are known for their intense radiation, strong stellar winds, and sometimes the presence of circumstellar material.
These factors can occasionally lead to the emergence of emission lines (mainly hydrogen) in these stars' spectra.
Examples include luminous blue variables (LBVs) and Herbig Ae/Be stars, which encompass massive evolved stars and intermediate-mass pre-main-sequence stars, respectively.
Studying the \ha emission in these massive stars provides essential insights into their physical properties, evolutionary stages, and circumstellar environments.

In particular, the primary focus of this paper revolves around Classical Be stars (hereafter CBe stars), which earned their name due to the presence of hydrogen emission lines (particularly {\ha}) in their spectra. 
CBes are fast-rotating variable stars exhibiting mass ejection episodes from their less-bound equator and decretion disks governed by viscosity \citep{rivinius2013}.
The presence of \ha emission lines in CBes is not merely used to confirm their classification; from spectroscopic studies, the strength and variability of these lines can provide insights into the dynamics and evolution of the disk, as well as the effects of processes such as mass loss, binary interactions, density waves, etc.
{In principle, the term Be star can be applied to any star with a B spectral type and emission lines. However, even though this term was indeed a source of confusion in the past, more recently, a taxonomically unique nomenclature was developed that distinguishes CBes from other emission B stars, e.g., B[e] supergiants, Ae/Be stars \citep{Porter03}.}

It has been long recognized that the variability of these stars is remarkably diverse, encompassing a wide range of amplitudes and time scales.
For instance, short-term, low-amplitude periodicity from non-radial pulsations are intrinsic to these objects (e.g., \citealt{rivinius2003}), and the disk build-up and dissipation phases can extend from days -- with outbursts producing detectable flickers \citep{labadie-bartz2020} -- to many years, such as $\kappa$~Draconis \citep{klement2022}, while ranging from tens of milimagnitudes to more than one magnitude in the visible range. Some systems have also shown periodicity in their long-term variability, such as  $\omega$~CMa \citep{ghoreyshi2021} and HD\,6226 \citep{2021MNRAS.508.2002R}, while others appear remarkably stable, such as $\beta$~CMi (see, e.g., Fig.~B.1 from \citealt{beatlas2023}). However, the prevailing scenario is likely unpredictable, leading to intricate and varied light curves.  \citep[e.g.,][]{2018AJ....155...53L}.

Fast rotation is a fundamental ingredient for the existence of CBe stars. To date, there are three proposed scenarios to explain the presence of such extreme rotators:
$i)$ CBe stars are born as fast rotators \citep{Bodenheimer71};
$ii)$ CBe stars are spun up by mass and angular momentum transfer from a binary companion (e.g., \citealt{pols1991}, \citealt{shao2014}); and
$iii)$ CBe stars begin their life as slow to moderate rotators, increasing their surface rotation rates along the main sequence (MS) (e.g., \citealt{ekstrom2008}, \citealt{georgy2013}).
The first two scenarios imply that fast rotation is already present at or close to the Zero Age Main Sequence (ZAMS), while the third one favors a higher incidence of CBe stars towards the end of their MS evolution.  
It is important to note that these scenarios are not mutually exclusive, and a combination of these processes could contribute to the formation of CBe stars.
Discriminating the relative prevalence of each scenario is a key step toward a fuller understanding of the stellar evolution of massive stars, particularly on the impact of binarity/multiplicity effects on the stellar evolution.

Several works have attempted to tackle this problem by studying CBe stars in young open clusters (YOCs) {using ground-based observations}.
Studying the ratio of CBe stars to the total B star population -- referred to as the Be/(B+Be) fraction -- in clusters of varying ages is a valid approach to unraveling the formation mechanisms of rapidly rotating stars; however, so far, methodological issues have hampered definite conclusions.
One complication is that different works employ different methodologies, making comparisons between results difficult and often leading to conflicting conclusions. 
For instance, \citet{Keller99} found that the incidence of CBe stars increases toward the MS turnoff, favoring the evolutionary channel, while \citet{martayan2006a} found several CBe stars of lower masses close to the ZAMS, which favors either the “born as fast rotators” or “binary evolution” scenarios. 
Additionally, many authors do not adequately consider significant effects that impact CBe star observations, such as gravity darkening and weak \ha emitters.
The last factor is often associated with tenuous disks that can lead to SED with H$\alpha$ emission that fills only partially the absorption profile. These CBe stars with weak disks are quite common \citep[e.g.][]{2017MNRAS.464.3071V} and often difficult to account for as CBes.

{\citet{milone2018} was the first study to effectively mitigate the resolution challenges of previous seeing-limited, ground-based observations. Their analysis was based on high-angular resolution and well-calibrated Hubble Space Telescope (HST) photometry of different young clusters in the SMC and LMC, including observations in the {\ha} filter with sufficient depth to sample {\ha} emitters corresponding to mid-to-late B-type stars.}

Previous studies of CBe stars in stellar clusters estimated the Be/(B+Be) fraction as a function of metallicity ($Z$) (e.g., \citealt{martayan2006a}, \citealt{martayan2007}) showing evidence that this number grows from about a tenth in the Galaxy to $\approx35\%$ in the Large Magellanic Cloud (LMC) and can be even higher in the Small Magellanic Cloud (SMC) \citep{martayan2010}, suggesting a strong $Z$-dependency of the Be phenomenon.

This paper is the first of a planned series to present consistent {ground-based} observations of YOCs with different ages and metallicities. This first publication aims to describe the technical details of data observation, reduction, and calibration pipelines while also showing the ability of adaptive optics imaging at the 4-m class SOAR telescope to resolve dense cores and provide photometric completeness for the entire B spectral type in the SMC. Furthermore, a preliminary analysis of the YOC NGC\,330 is presented.

The paper is structured as follows. Section~\ref{sec:ngc330} presents a brief review of NGC\,330 and its significance in CBe research.
Section~\ref{sec_obs} describes the observations and subsequent reduction procedures.
As follows, Sect.~\ref{sec_photcal} details the photometric calibration procedure and the comparison with external catalogs.
Finally, the photometric results and initial analysis of the NGC\,330 observations are presented in Sect.~\ref{sec_results}, and the main conclusions are discussed in Sect.~\ref{sec_discussion}.

\section{\texorpdfstring{NGC\,330}{NGC330}}
\label{sec:ngc330}

NGC\,330 is a young open cluster located at $\alpha = 00$h\,$56$m\,$17.6$s and $\delta = -72^{\circ}\,27\arcmin\,47\arcsec$ (J2000, \citealt{ngc330coord}), within the Small Magellanic Cloud (SMC). NGC\,330 was chosen for this pilot study due to its relatively large CBe content. It was the subject of intense scrutiny in the last decades, with some stars being photometrically documented over sixty years ago in a survey conducted by \citet{smc1}. The first detection of CBes in NGC\,330 was done by \citet{smc2} by spectroscopically measuring the equivalent width of the \ha line ($\lambda$=6562.8~{\AA}). 
Emission line stars, particularly active CBe stars, can be identified using color index information based on the narrow-band \ha filter and another filter encompassing the nearby continuum. 
For NGC\,330, such methodology was first introduced by \citet{grebel1992}, providing a significant breakthrough in detecting CBes associated with the YOC. Their Fig.~2 summarizes the adopted classification method through a (H$\alpha-y$, $b-y$) color-color diagram constructed from broad-band Strömgren $by$ filters and a custom narrow-band \ha filter, where active CBes with sufficiently dense disks deviate from the expected MS location.
Later, \citet{smc10} validated the methodology adopted by \citeauthor{grebel1992}, observing NGC\,330 using $\mli{VRI}$+{\ha} filters.

NGC\,330 observations have provided substantial support for various subsequent discoveries. These include a detailed analysis of blue stragglers and evolved stars, comparing photometry with the available evolutionary models at that time \citep{smc14}. Additionally, the application of photometric separation between CBes and normal B-type stars in clusters from the LMC and SMC, including NGC\,330, was further explored in works by \citet{Keller99} and \citet{Iqbal13}. Other studies delved into the relationship between CBe incidence and metallicity, as discussed in \citet{martayan2006a} and \citet{martayan2007}. Moreover, investigations on stellar variability using OGLE II data were conducted by \citet{smc21}. More recently, \citep{Bodensteiner20} reported the first results from AO-assisted integral-field MUSE spectroscopic observations of the densest region of NGC\,330, resolving and classifying a flux-limited sample of B and Be stars and providing an age estimate of the cluster in range of 35 to 40~Myr. At the same time, \citet{smc24} reported an independent assessment of the age of NGC\,330 as 45$\pm$5~Myr based on the analysis of the evolved red supergiant population of the cluster.

\section{Observations and Data Reduction}
\label{sec_obs}

Kron-Cousins $\mli{BVRI}$+{\ha} imaging observations were carried out at the 4.1-m SOAR telescope (Chile) on January 5th, 2019. We used the SOAR Adaptive Module \citep[SAM, ][]{Tokovinin16}, a GLAO system using a Rayleigh laser guide star at $\sim$7~km from the telescope to correct the effects of ground-layer atmospheric turbulence.
SAM provides corrected images to its internal CCD detector, SAMI (4k\,$\times$\,4k~CCD), operating in a 2\,$\times$\,2 binned mode, resulting in a plate scale of 91~mas\,pixel$^{-1}$ over a field-of-view (FOV) of $\sim$3\farcm1\,$\times$\,3\farcm1 on the sky.

Individual, long-exposure $\mli{BVRI}$ images were taken using a three-point dither pattern, separated by {15\arcsec} in the E–W direction, with exposure times of 60~s per frame, totaling 180~s per filter. For {\ha}, three individual exposures of 240~s were obtained (totaling  12~min).
Observations included short exposures to avoid saturation of the brightest stars in the FOV. The short exposures were taken in a two-point dither pattern separated by {15\arcsec} in the E–W direction, with exposure times of 5~s per frame, totaling 10~s per filter.
Table~\ref{table_log} summarizes the SAMI observations.

\setlength{\tabcolsep}{1.85pt}
\begin{table}[ht]
    \centering
    \caption{Log of the SAMI observations for NGC\,330 made on January 5th, 2019.}
    \label{table_log}
    \begin{tabular}{ccccccc}
    \hline
    \hline
        Filter            & MJD             & $\mathrm{N_{exp}}$ & $t_{\mathrm{exp}}$ & Airmass & Atm. Seeing  \\
                   &                 &      & (s)    &         & (\arcsec)   \\
    \hline
        $B$	& 58488.05261(50)	&  3 & 60	& 1.47	& 0.54 \\
        $B$	& 58488.05462(06)	&  2 & 5	& 1.47	& 0.53 \\
        $V$	& 58488.04946(50)	&  3 & 60	& 1.46	& 0.50 \\
        $V$	& 58488.04823(06)	&  2 & 5	& 1.46	& 0.47 \\
        $R$	& 58488.05661(50)	&  3 & 60	& 1.47	& 0.55 \\
        $R$	& 58488.05536(06)	&  2 & 5	& 1.47	& 0.51 \\
        $I$	& 58488.05968(50)	&  3 & 60	& 1.48	& 0.49 \\
        $I$	& 58488.06229(07)	&  2 & 5	& 1.49	& 0.50 \\
    {\ha}	& 58488.06650(19)	&  3 &240	& 1.50	& 0.54 \\
    {\ha}	& 58488.06316(06)	&  2 &  5	& 1.49	& 0.59 \\
    \hline
    \end{tabular} \\ \vspace{0.1cm}
    \noindent \justifying{{Notes:} The mean MJD of the observations are provided, and the errors on the last two decimal places are indicated within the parenthesis. The mean atmospheric seeing was obtained from the DIMM monitor installed at Cerro Pach\'on.}
\end{table}
\setlength{\tabcolsep}{6pt}

\subsection{Data Reduction}

The data was processed using the SAMI PyRAF-based PySOAR pipeline\footnote{\url{https://noirlab.edu/science/programs/ctio/instruments/sam/user-guides/reducing-your-sam-images}.}.
Initially, the pipeline creates master bias and master flat-field templates for each filter.
Subsequently, the raw images are corrected for overscan, bias-subtracted, and flat-fielded. 

The astrometric solution and distortion correction of the field were obtained by comparing the field with reference stars in the USNO-A2 catalog using \texttt{SExtractor} and \texttt{Scamp} \citep{Bertin02}, leading to positional errors smaller than 0\farcs15. Finally, the individual images were average-combined using \texttt{Swarp} \citep{Bertin02}.

\begin{figure*}[!ht]
\centering
{\includegraphics[viewport=50bp 535bp 820bp 900bp, clip, width=\textwidth]{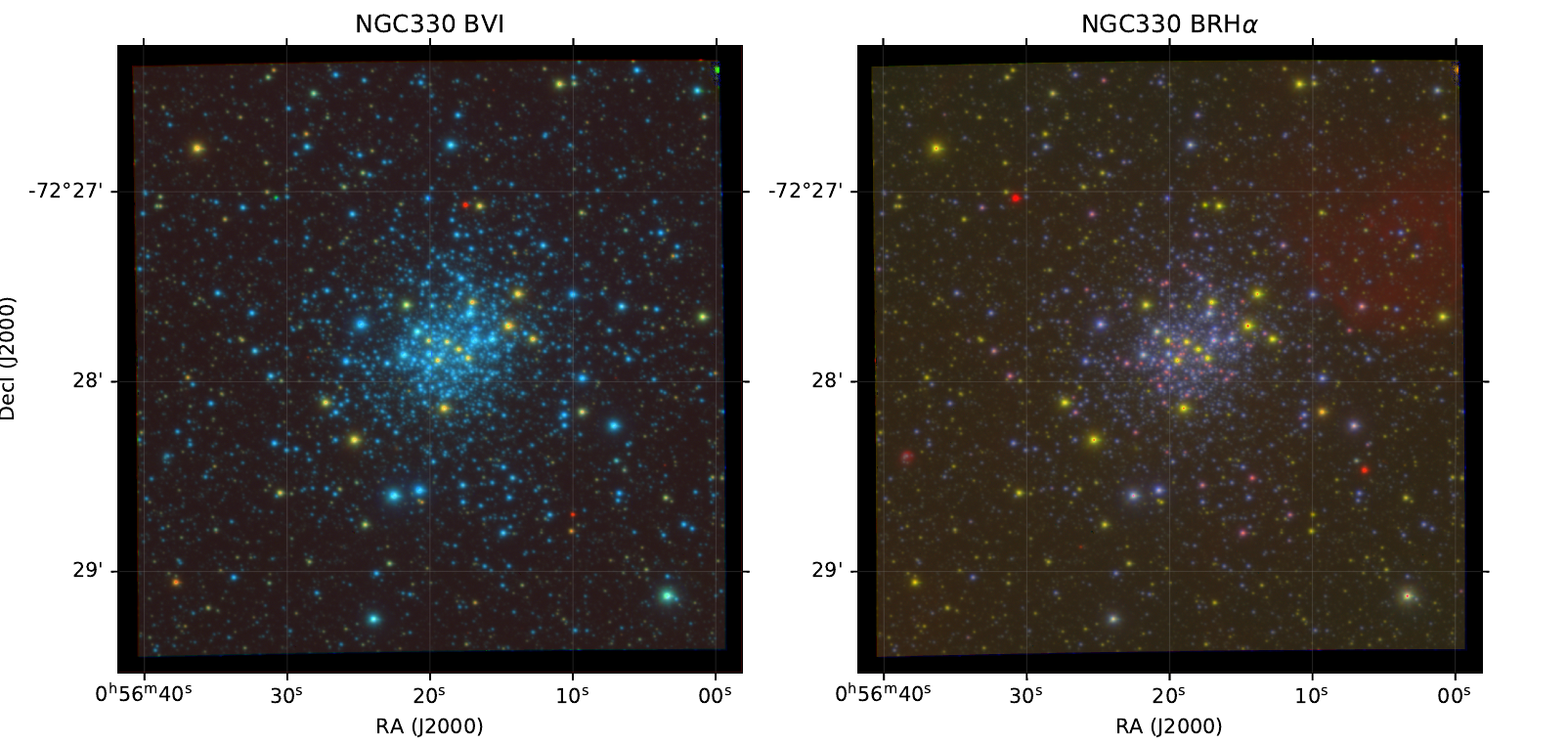}}\\[-1.5ex]
\caption{False-color SAMI images of NGC\,330 based on the long-exposure frames. 
Left: $\mli{BVI}$ image (blue: $B$, green: $V$, red: $I$). Right:  $BR$H$\alpha$ image  (blue: $B$, green: $R$, red: H$\alpha$). Both images have a 3\farcm1\,$\times$\,3\farcm1 field-of-view, centered at
$\alpha = 00$h\,$56$m\,$20.0$\,s and $\delta = -72^\circ \,27\arcmin\,51\farcs67$
(J2000). 
North is up, and east is to the left.}
\label{fig_rgb}
\end{figure*}

Figure~\ref{fig_rgb} presents false-color RGB maps using the long-exposure $\mli{BVRI}$+H$\alpha$ images. The left panel shows the $\mli{BVI}$ images, exhibiting a few reddened sources in the field. The right panel shows the $\mli{BR}$+{\ha} images. It is possible to identify an extended H\,{\sc{ii}}-like structure to the NW direction of NGC\,330, and very red sources associated with a strong H$\alpha$ emission. 

\subsection{PSF photometry}

We extracted the stellar photometry using an automated PSF fitting procedure based on the IDL StarFinder algorithm \citep{Diolaiti00}.

\begin{figure}[!ht]
\centering
\includegraphics[viewport=60bp 360bp 450bp 555bp, clip, width=\columnwidth]{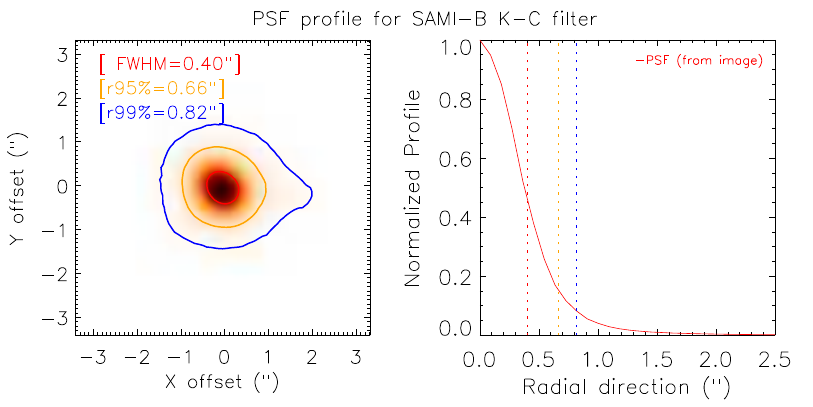}
\includegraphics[viewport=60bp 360bp 450bp 555bp, clip, width=\columnwidth]{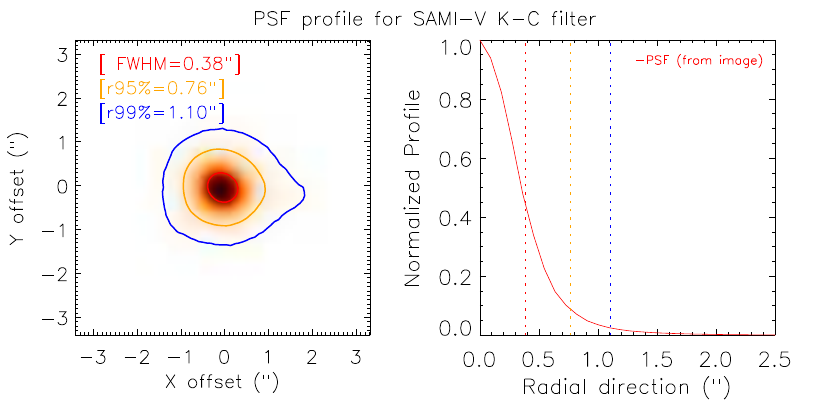}
\includegraphics[viewport=60bp 360bp 450bp 555bp, clip, width=\columnwidth]{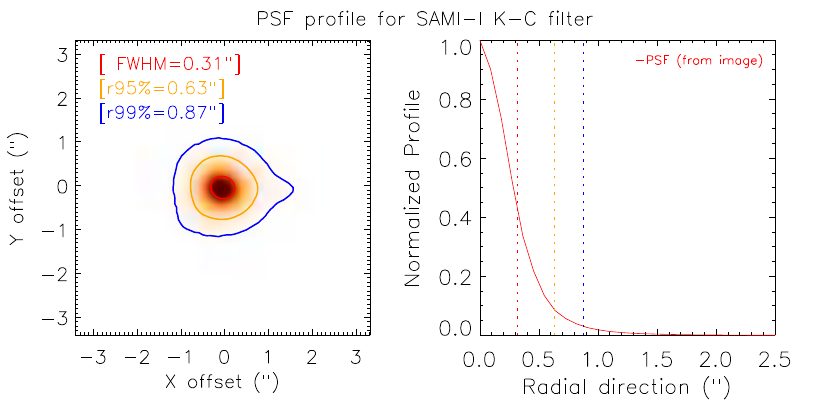}\\[-1.5ex]
\caption{PSF models for the SAMI $\mli{BVI}$-band images for NGC\,330.
\textit{Left panel:} two-dimensional profile of the PSF overlaid by the FWHM (red contour), and the radii containing 95\% (orange) and 99\% (blue) of the enclosed fluxes, respectively.
\textit{Right:} radial profile of the PSF. The vertical dashed lines indicate the FWHM (red), and the radii containing 95\% (orange) and 99\% (blue) of the enclosed flux.}
\label{fig_PSFs}
\end{figure}

We used relatively isolated and bright sources to construct the PSF of each $\mli{BVRI}$+{\ha} image. Examples of the resulting PSF template for the $\mli{BVI}$ images are presented in Fig.~\ref{fig_PSFs}. The PSF models exhibit an extended weak tail towards positive $x$-axis offsets, a common consequence of the divergence of the tip-tilt loop, according to the instrument manual\footnote{SAM Operator's Manual in \url{https://noirlab.edu/science/programs/ctio/instruments/sam/support-staff}.}.

Table~\ref{table_psf_fitting} summarizes the results from the PSF fitting for each filter, showing the extraordinary performance of the AO-assisted observations, achieving angular resolutions as sharp as 0\farcs33 in the {\ha} filter.

\setlength{\tabcolsep}{1.8pt}
\begin{table}[htb]
    \centering
    \caption{Parameters of the PSF model for SAMI observations and the peak and limiting magnitude for each filter.}
    \label{table_psf_fitting}
    \begin{tabular}{c|ccccccc} 
    \hline
        Filter & FWHM      & $N_{\ast}$  & $N_{\ast}$ & 1-$\sigma$ & $m_{\mathrm{peak}}$ & $m_{\mathrm{lim}}$ & $\sigma$ \\
               & (\arcsec) & {\footnotesize (short)} & {\footnotesize (long)} & {\footnotesize (ADU)} & {\footnotesize (mag)}  & {\footnotesize (mag)} & {\footnotesize (mag)} \\
    \hline
        B  & 0.40 & 1262 &  4820 & 1.86  & 21.3  & 22.5 & 0.06-0.14 \\
        V  & 0.38 & 2101 &  7057 & 2.19  & 21.7  & 23.4 & 0.03-0.18 \\
        R  & 0.35 & 3255 &  9464 & 3.40  & 22.2  & 24.2 & 0.03-0.30 \\
        I  & 0.31 & 1840 &  7876 & 3.58  & 21.7  & 22.8 & 0.04-0.26 \\
        H$\alpha$ & 0.33 &  424 &  5715 & 2.05  & 22.0  & 23.6 & 0.09-0.21 \\
    \hline
    \end{tabular} \\ \vspace{0.1cm}
    \noindent \justifying{{Notes:} The FWHM was measured for the PSF of each filter. $N_{\ast}$ corresponds to the number of sources detected above a 1-$\sigma$ limit in the short and long exposure frames. $\sigma$ corresponds to the range of magnitude uncertainties for each filter.}
\end{table}
\setlength{\tabcolsep}{6pt}

The identification and extraction of the fluxes of the point-like sources were performed by adopting a total of three iterations, considering 3-, 2-, and 1-$\sigma$ thresholds, and a list of positions ($x,y$) and instrumental fluxes (in ADU) was obtained for each image. The ($x,y$) positions were transformed into sky coordinates (RA and Dec) using the astrometric information stored in the header of the images. The instrumental fluxes were corrected to the exposure time of 1~s, and instrumental magnitudes $bvri$+h$\alpha$\footnote{By definition, lower case and capital letters denote instrumental and calibrated magnitudes, respectively.} were evaluated using a zero point of 25.0~mag.

\begin{figure}[!ht]
\centering
\includegraphics[viewport=10bp 10bp 600bp 390bp, clip, width=\columnwidth]{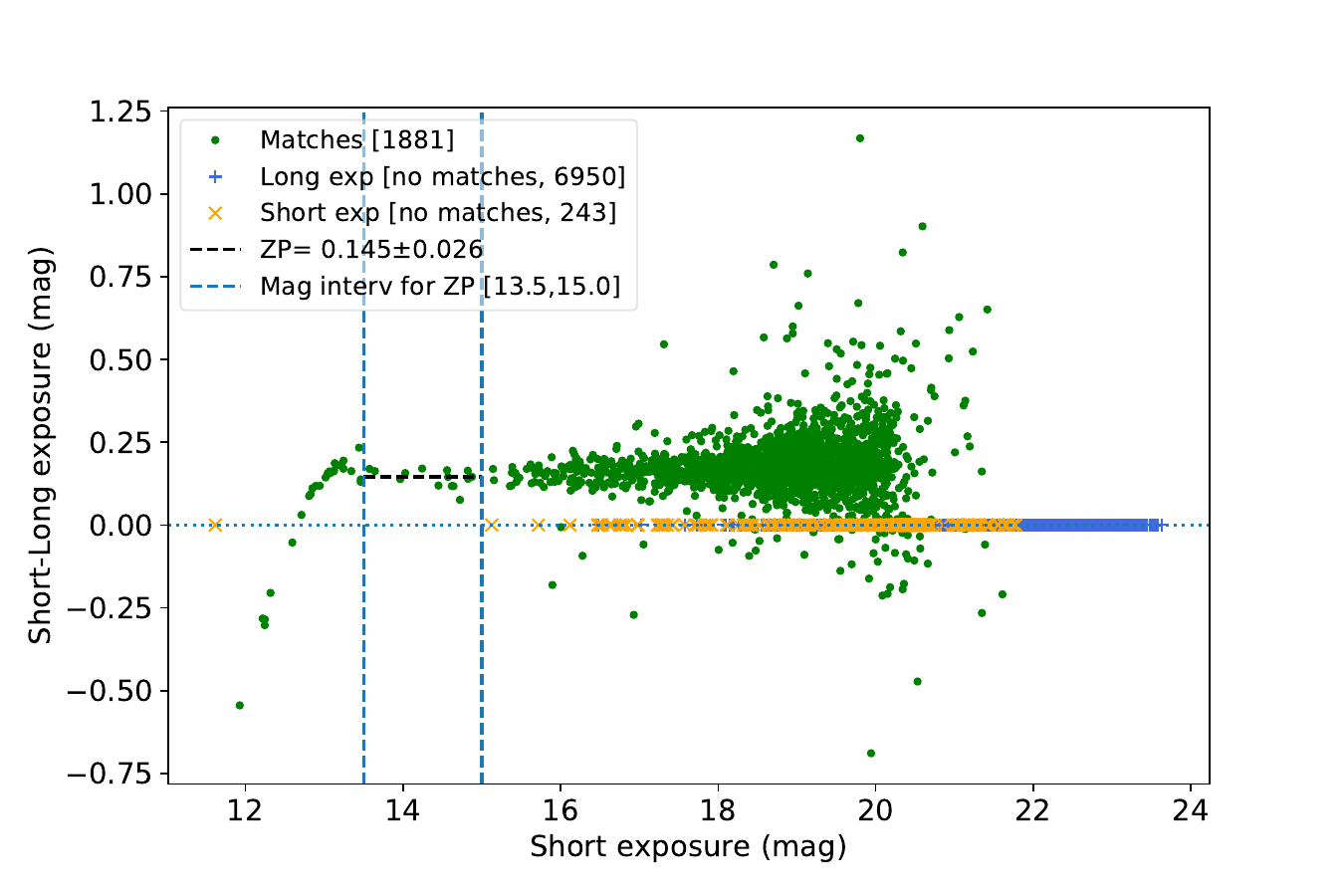}
\caption{Distribution of residuals between the instrumental magnitudes from short- and long-exposure images for the $I$-band filter. The zero-point offset between short- and long-exposure magnitudes (ZP=0.145$\pm$0.026~mag, indicated by the horizontal dashed black line) was estimated from the points located within the vertical dashed blue lines. Saturated sources ($\lesssim 13.5$~mag) were retrieved from the short-exposure list only.
\label{fig_merge_mags}}
\end{figure}

Figure~\ref{fig_merge_mags} presents the residuals between the magnitudes extracted from short and long exposure lists for the $I$-band.
The short and long exposure lists were merged, and a final catalog per filter was created as follows.
We evaluated a relative zero point (ZP) between the magnitudes from the short and long lists, that is
\[
  m_{\lambda,\rm short} = m_{\lambda,\rm long} + \mathrm{ZP}_{\lambda}    
\]
\noindent where the ZP per filter are: $\mathrm{ZP}_b\,=\,0.036(12)$\,mag, $\mathrm{ZP}_v\,=\,0.015(12)$\,mag, $\mathrm{ZP}_r\,=\,0.063(23)$\,mag, $\mathrm{ZP}_i\,=\,0.146(26)$\,mag, and $\mathrm{ZP}_{h\alpha}\,=\,-0.094(25)$\,mag.
The ZPs were estimated from the data points that are relatively brighter but not saturated in each filter, as indicated by the vertical dashed blue lines in Fig.~\ref{fig_merge_mags}. 
Sources saturated on the long-exposure images (corresponding to instrumental magnitudes brighter than 12.0, 13.5, 13.4, 13.7, and 13.7~mag in $\mli{BVRI}$+{\ha} filters, respectively) were retrieved from the short-exposure list. Fainter sources were obtained from the long-exposure list.
We adopted the mean instrumental magnitude value for all the other detections identified in both lists.
Finally, the individual filters were cross-matched one by one, assuming a maximum separation of 0\farcs3,
resulting in a final list of 10220 entries with detections in at least one filter, and 4430 entries with magnitudes detected in all five filters. 

\section{SAMI photometry}
\label{sec_photcal}

\subsection{Photometric calibration}
\label{sec_cal_templatefitting}

We implemented a \emph{stellar template fitting} methodology for the photometric calibration of the SAMI $\mli{BVRI}$+{\ha} photometry, following a similar procedure adopted for the \emph{Southern Photometric Local Universe Survey} \citep[S-PLUS,][]{splusdr2}. The method consists of calibrating the instrumental magnitudes using a library of synthetic photometry, calculated with a grid of atmospheric stellar models and an external photometric reference catalog covering the same region of the sky. The complete calibration process was developed in Python, using functions from the \texttt{AstroPy} and \texttt{SciPy} libraries.
The following sequence outlines the process:

\begin{enumerate}

\item {A list of sources with available magnitudes in the external and the SAMI catalogs is used as input. The external reference catalog serves as a source of pseudo-standard stars.}

\item {The photometric spectral energy distribution (SED) of every star from the list is fitted using the synthetic library, finding a representative model. As atmospheric models provide surface fluxes instead of absolute fluxes, the fitting is performed using a set of color indices (e.g., $g-r$, $r-z$, etc.);
}

\item A mean magnitude difference (representing every filter from the external catalog) between the best model and the {corresponding} star is determined; 

\item The SAMI magnitudes are then predicted using the stellar model. The residuals between the predicted and instrumental magnitudes are used to infer a photometric ZP for every filter;

\item A final ZP of every filter is estimated based on the residuals obtained in item~(4), as presented in Fig.~\ref{fig_cal_ext_zp},
and applied for every star in the SAMI catalog.

\end{enumerate}

One of the main advantages of this approach is the avoidance of transformation equations (especially for narrow-band filters). Moreover, it treats extinction as a separate and independent parameter, mapped by convolving the synthetic spectra with a range of $R_V$ and $E(B-V)$ values.

In this method, the ZP system inevitably aligns with the external catalog. {Therefore, the spectral coverage of the external catalog must be similar to or broader than that of the observations, as a necessary condition for obtaining a precise photometric calibration.}

%
We used the synthetic spectral library of \citet{Coelho14}, which provides model SEDs ranging from 0.13 to 10~{\micron}, with resolutions of $\Delta\log\lambda=8\times10^{-4}$ and a wide range of stellar parameters: effective temperature (\teff) from 3,000 to 25,000~K, surface gravity from $\log(g) = -0.5$ to $+5.5$ (in steps of 0.5~dex), and metallicity ($Z$) from $Z=0.0017$ to 0.049 in a custom set of 12 chemical mixtures (for more details, see Table~1 of \citealt{Coelho14}).

%
We adopted the SkyMapper\,DR1.1 \citep{skymapper} as the external reference catalog, combining an excellent photometric coverage of the southern hemisphere in the SLOAN/SDSS $uvgriz$ filters, providing magnitudes in the AB system and covering a broader spectral range than our observations.

We cross-matched the SAMI coordinates with SkyMapper, using a maximum radius threshold of 0\farcs3 between the catalogs{. This resulted} in a total of 204 common sources in both catalogs, out of which 82 exhibit magnitudes detected in all six $uvgriz$ filters.

%

The library of synthetic photometry was constructed as follows. 
Each spectrum of the atmospheric library was reddened by following the \citet{cardelli1989} extinction law with a fixed value of $R_V=3.1$, and assuming 22 different $E(B-V)$ values ranging between 0 and 1.0~mag. 
Next, the reddened spectra were convolved with the transmission curves of the filters from both the external catalog and the SAMI sets. This process generated a list of 11 magnitudes ($\mli{BVRI}+$H$\alpha$ + $uvgriz$) for each synthetic spectrum.

The application of the \emph{stellar template fitting} procedure is divided into external and internal calibration steps, explained as follows.


\subsubsection{External calibration}
\label{sec_cal_ext}

The external calibration procedure compares the external catalog with each atmospheric model by computing eight color indices based on the SkyMapper filter set (e.g., $u-v$, $g-i$, etc).
The best match between the synthetic library and the catalog is selected through a $\chi^2$ minimization process, where the $\chi^2$ value is computed from the contribution of every color index ($C_i$) in the form:
\begin{equation}
\chi^2 = \sum_{\mathrm{i=1}}^{8}\left(\frac{\mathrm{C_{i,model}} - \mathrm{C_{i,ext}}}{\mathrm{e_{C_{i,ext}}}}\right)^2,
\label{eq_chi2}
\end{equation}
\noindent where 
$\mathrm{e_C}$ is the uncertainty of the color index given by $\mathrm{e_C} = (\mathrm{C1}^2 + \mathrm{C2}^2)^{1/2}$, where $\mathrm{C1}$ and $\mathrm{C2}$ represent the magnitudes of $\mathrm{e_C}$.
The best-fitting model configuration for a specific source is determined by the minimum value of $\chi^2$ among all possible models considered for that source.

To filter out unrealistic matches between the external catalog and the models, we introduced a parameter termed the \emph{mean photometric offset} ($\mu_{\Delta}$) and its corresponding error ($\sigma_{\mu_\Delta}$), explained as follows.
For each target, we computed the difference between the magnitude from the external catalog ($m_\mathrm{ext}$) and the one from the best-fitted model ($m_\mathrm{best\,model}$) as:
\begin{equation}
\Delta m_i = m_{i,\mathrm{ext}} - m_{i,\mathrm{best\,model}} \; \, \mathrm{for} \, \, i \in uvgriz\,, 
\label{eq_deltamag}
\end{equation}
\noindent Following the definition of  $\mu_{\Delta}$  and $\sigma_{\mu_\Delta}$ as the mean and standard deviation of $\Delta m_i$ values across all six $uvgriz$ filters, we established that a suitable fit should display consistent $\Delta m_i$ values for all filters. Consequently, $\sigma_{\mu_\Delta}$ should approach zero. Hence, we only regarded fits as satisfactory for sources exhibiting $\sigma_{\mu_\Delta} \leq 0.15$~mag.

The predicted magnitudes for each source, derived from the best model magnitude value in the external catalog and applied to the SAMI filters ($m_\mathrm{pred}$) were evaluated as:
\begin{equation}
m_{i,\mathrm{pred}} = m_{i,\mathrm{best\,model}} + \mu_{\Delta} \; \, \mathrm{for} \, \, i \in \mli{BVRI}\texttt{+}\mathrm{H}\alpha
\label{eq:externalpredicted}
\end{equation}

\begin{figure*}[ht!]
\centering
{\includegraphics[viewport=10bp 25bp 1440bp 350bp, clip, width=\textwidth]{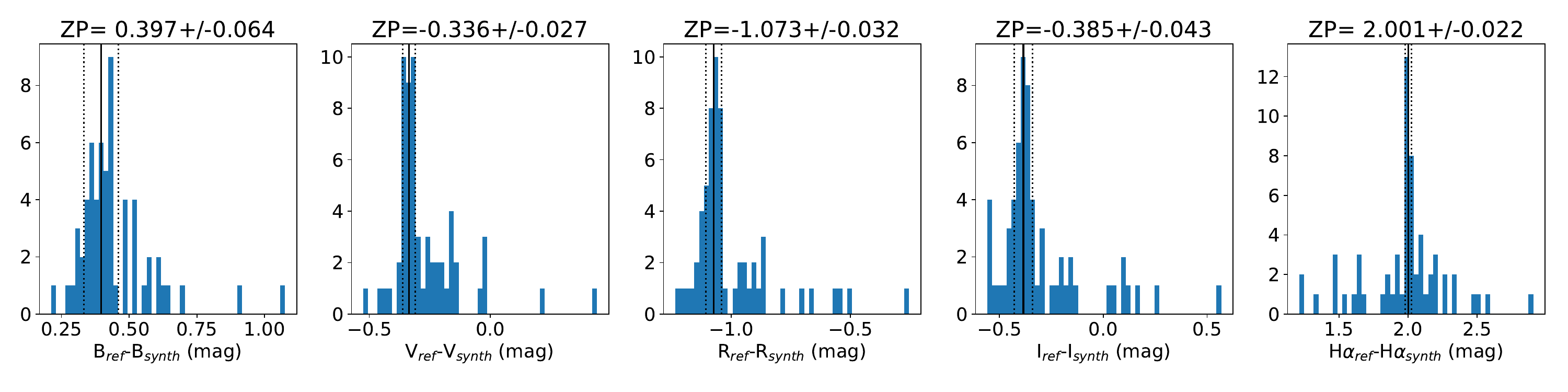}} \\[-1.0ex]
\caption{Distribution of residuals between the observed and synthetic model magnitudes of the external reference catalog. The external photometric zero-point (ZP) was estimated through an outlier-resistant Gaussian fitting procedure. The ZP and the 1-$\sigma$ error bars (indicated at the title of each plot) are shown as the filled and dashed vertical lines, respectively.
\label{fig_cal_ext_zp}}
\end{figure*}

The photometric zero point offset for each filter ($\mathrm{ZP}_{\mathrm{ext},i}$) was inferred from the residuals between the instrumental magnitude ($m_{\mathrm{inst}}$) and the predicted magnitude ($m_\mathrm{pred}$) of the sources from the external catalog:
\begin{equation}
\mathrm{ZP}_{\mathrm{ext},i} = m_{i,\mathrm{inst}} - m_{i,\mathrm{pred}} \; \, \mathrm{for} \, \, i \in \mli{BVRI}\texttt{+}\mathrm{H}\alpha.
\label{eq_zp}
\end{equation}
\noindent The outcome of Eq.~(\ref{eq_zp}) is a distribution of ZP values, as presented in Fig.~\ref{fig_cal_ext_zp}. For each filter, the distribution exhibits a defined peak containing the bulk of the sources and an extended tail of outliers toward positive values. An outlier-resistant Gaussian fit was performed on each histogram, and the mean external ZP and its error ($\sigma_{\mathrm{ZP}}$) correspond to the Gaussian $\mu$ and $\sigma$, respectively. Given that the fitting procedure excludes outliers, the associated error is relatively low (e.g., see the last plot for the {\ha} filter).

Finally, the calibrated SAMI magnitudes ($m_{\rm ext}$) and their errors ($\sigma_{\rm ext}$) were obtained by applying the fitted ZPs as:
\begin{equation}
\left.
\begin{array}{cc}
m_{i,\mathrm{ext}} & = m_{i,\mathrm{inst}} - \mathrm{ZP}_{i} \\
\sigma_{i,\mathrm{ext}} & = \sqrt{\sigma_{i,\mathrm{inst}}^2 + \sigma_{\mathrm{ZP}_i}^2}
\label{eq:externalcalibfinal}
\end{array}
\right\} \; \, \mathrm{for} \, \, i \in \mli{BVRI}\texttt{+}\mathrm{H}\alpha,
\end{equation}

\noindent In principle, Eq.~(\ref{eq:externalcalibfinal}) yields suitable calibrated magnitudes for scientific purposes. 
However, the photometric calibration can be improved through an internal calibration step, as explained below.


\subsubsection{Internal calibration}
\label{sec_cal_int}

The internal calibration is very similar to the previous procedure but uses the entire externally calibrated SAMI catalog as input, replacing the external reference catalog. As a consequence, this step is significantly time-consuming.
%
{However, the internal calibration provides two important byproducts: $i)$ Validation of the external calibration procedure for SAMI magnitudes and improvement in photometric calibration, particularly for filters without a direct counterpart in the external catalog (e.g. narrow-band filters); and
$ii)$ Obtaining extinction and atmospheric parameters from the best-fitted model for all sources in the externally calibrated photometric catalog.}
{Instances where internal calibration improves photometric results are typically linked to multi-band observations utilizing several narrow-band filters, such as those in the S-PLUS survey \citep{splusdr2}. In addition, the second byproduct provides a complementary and independent result from the improvement of the photometry, which can be used in a broader range of applications.}

In short, each source is matched against the synthetic magnitude library, and a $\chi^2$ minimization procedure is obtained for five color indices based on the broad-band $\mli{BVRI}$ filters, 
following a similar approach as outlined in Eq.~(\ref{eq_chi2}).
The \emph{internal mean photometric offset} ($\mu_{\Delta,\mathrm{int}}$) and its error ($\sigma_{\mu_{\Delta,\mathrm{int}}}$) were evaluated in the same way as the external calibration procedure.
In this case, $\sigma_{\mu_{\Delta,\mathrm{int}}}$ serves as an indicator of the quality of fit between the calibrated magnitudes and their corresponding counterparts in the synthetic photometric library: values closer to zero signify well-fitted models, while larger values indicate poorer fits.
Then, the internally-calibrated magnitudes ($m_\mathrm{pred}$) of the source are calculated as:
\begin{equation}
m_{i,\mathrm{pred}} = m_{i,\mathrm{best\,model}} + \mu_{\Delta,\mathrm{int}} \; \, \mathrm{for} \, \, i \in \mli{BVRI}\texttt{+}\mathrm{H}\alpha.
\label{eq:internalpredicted}
\end{equation}

\begin{figure*}[ht!]
\centering
{\includegraphics[viewport=15bp 25bp 1440bp 350bp, clip, width=\textwidth]{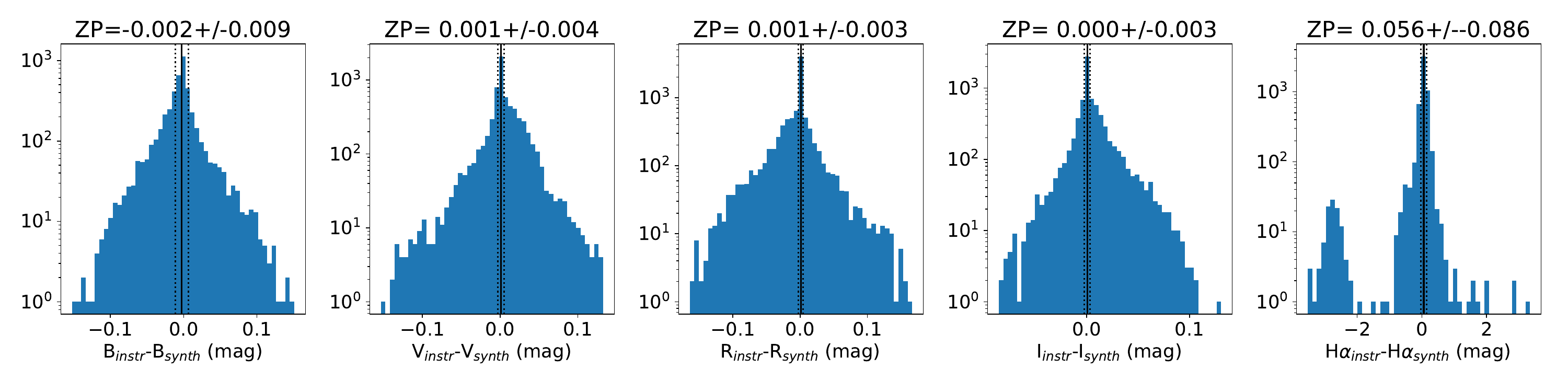}} \\[-1.0ex]
\caption{Distribution of residuals between the observed and synthetic model magnitudes of the SAMI magnitudes (internal calibration). The $y$-axis is shown on a logarithmic scale. A full description of the plots is given in Fig.~\ref{fig_cal_ext_zp}.
\label{fig_cal_int_zp}}
\end{figure*}

Similarly to the external calibration method, an internal zero-point offset ($\mathrm{ZP}_{\mathrm{int}}$) can be evaluated from the distribution of the residuals between $m_{i,\mathrm{inst}}$ and $m_{i,\mathrm{pred}}$, as presented in Fig.~\ref{fig_cal_int_zp}.
For successful calibrations, all $\mathrm{ZP}_{\mathrm{int}}$ values are expected to be close to zero, and therefore, no internal calibration is necessary. For cases where the ZP of a filter is larger than its error, the internal calibration process is a necessary step, providing a better constraint on that region of the SED. In such cases, the internal ZPs need to be accounted for in the calibration process as follows:
\begin{equation}
\left.
\begin{array}{cc}
m_{i,\mathrm{int}} & = m_{i,\mathrm{ext}} - \mathrm{ZP}_{\mathrm{int},i} \\
e_{i,\mathrm{int}} & = \sqrt{e_{i,\mathrm{ext}}^2 + e_{i,\mathrm{ZP}_\mathrm{int}}^2}
\label{eq:internalcalibfinal}
\end{array}
\right\} \; \, \mathrm{for} \, \, i \in \mli{BVRI}\texttt{+}\mathrm{H}\alpha.
\end{equation}

Indeed, Fig.~\ref{fig_cal_int_zp} illustrates that histograms of the broad-band $\mli{BVRI}$ filters are centered around zero with relatively small errors ($\lesssim 0.01$~mag).
This observation signifies the successful application of the photometric calibration to the SAMI photometry. 
{As previously mentioned, the internal calibration does not improve the overall photometric results presented in this work. However, its extinction and stellar parameter byproducts are presented as a relevant result for further works.}

Unlike the other filters, the zero point of the {\ha} filter exhibits both the largest deviation from zero (ZP\,=\,0.056~mag) and the largest error ($\sigma_\mathrm{ZP}=\pm0.083$~mag).
Given that the {\ha} filter serves as an indicator of stellar activity processes (including disk emission from CBe stars), we do not expect it to have the same dispersion observed in the broad-band $\mli{BVRI}$ filters.
In fact, the {\ha} residuals exhibit a bimodal distribution where most of the sample is centered at zero, and a smaller fraction is concentrated around $-3$~mag. {Such distribution} indicates the bulk of {\ha} fluxes is consistent with the synthetic models, as expected for normal stars, but a significant number of sources exhibit \ha fluxes larger than predicted by the synthetic models. This is an important output from the internal calibration method, allowing us to identify the sources with \ha excess. 

\subsection{Calibrated SAMI magnitudes}

Table~\ref{table_phot_SAMI} presents the calibrated SAMI photometric results obtained from the external calibration procedure (Sect.~\ref{sec_cal_ext}), together with the stellar parameters from \citet{Coelho14}, delivered by the internal calibration (Sect.~\ref{sec_cal_int}). 
The uncertainties of the calibrated magnitudes range from 0.03 to 0.30~mag for the $\mli{BVRI}$ filters, and from 0.02 to 0.19~mag for the {\ha} filter.
The full list of the photometric results is available online.

\begin{table*}[!ht]
    \centering
    \setlength\tabcolsep{1.5pt}
    \caption{Calibrated $\mli{BVRI}+\mathrm{H}\alpha$ SAMI photometry for NGC\,330, the associated physical parameters from the internal calibration method and the photometric classification of the source.}
    \label{table_phot_SAMI}
    \begin{tabular}{rcccccccccrcrccc}
    \hline
    \hline
ID & RA & Dec & N$_\mathrm{filt}$ & $B$ & $V$ & $R$ & $I$ & $H\alpha$ & $E_{B-V}$ & $T_{\mathrm{eff}}$ & $\log(g)$ & Fe/H & $\alpha$ & $\sigma_{\mu_{\Delta,int}}$ & Type\\
{} & (deg) & (deg) &  & (mag) & (mag) & (mag) & (mag) & (mag) & (mag) & (K) & (dex) & (dex) & {} & (mag) & \\
    \hline
1 & 14.070568 & $-$72.463074 & 5 & 12.494(64) & 12.619(29) & 12.815(40) & 13.048(51) & 12.769(21) & 0.000 & 9250 & 2.0 & $-$0.5 & 0.4 & 0.0002 & MS\\
2 & 14.094029 & $-$72.476657 & 5 & 12.523(64) & 12.535(29) & 12.652(40) & 12.834(51) & 12.636(21) & 0.000 & 7750 & 2.0 & 0.0 & 0.0 & 0.0048 & MS\\
3 & 14.014281 & $-$72.485452 & 5 & 12.671(64) & 12.339(29) & 12.180(40) & 12.167(51) & 12.123(21) & 0.000 & 6250 & 4.5 & $-$1.0 & 0.4 & 0.0133 & MS\\
4 & 14.103593 & $-$72.461670 & 5 & 12.741(64) & 12.920(29) & 13.164(40) & 13.460(51) & 13.104(21) & 0.025 & 13000 & 4.0 & 0.0 & 0.0 & 0.0007 & MS\\
5 & 14.091106 & $-$72.464351 & 5 & 12.972(64) & 13.050(29) & 13.207(40) & 13.426(51) & 13.220(21) & 0.025 & 10000 & 1.5 & $-$1.0 & 0.4 & 0.0046 & MS\\
6 & 14.065419 & $-$72.462967 & 5 & 13.017(64) & 13.107(29) & 13.284(40) & 13.484(51) & 13.300(21) & 0.000 & 8500 & 1.5 & $-$1.0 & 0.0 & 0.0013 & MS\\
7 & 14.071939 & $-$72.460710 & 5 & 13.032(64) & 13.116(29) & 13.274(40) & 13.484(51) & 13.305(21) & 0.025 & 10000 & 1.5 & $-$1.0 & 0.4 & 0.0016 & MS\\
8 & 14.086628 & $-$72.476189 & 5 & 13.069(64) & 13.238(29) & 13.491(40) & 13.814(51) & 13.435(21) & 0.000 & 12250 & 4.5 & 0.2 & 0.4 & 0.0048 & MS\\
9 & 14.083651 & $-$72.464272 & 5 & 13.120(64) & 13.220(29) & 13.426(40) & 13.682(51) & 13.536(21) & 0.050 & 12250 & 4.5 & 0.0 & 0.0 & 0.0014 & MS\\
10 & 14.087173 & $-$72.462312 & 5 & 13.254(64) & 13.223(29) & 13.261(40) & 13.358(51) & 13.268(21) & 0.000 & 7000 & 1.0 & $-$1.0 & 0.0 & 0.0100 & MS\\
    \hline
    \end{tabular}%
    \noindent \justify{\footnotesize{Notes:} Only the first 10 rows of the table are shown. The full table is available online. 
The columns are:
(1) Index
(2--3) Right ascension and declination (in degrees, J2000);
{(4) Number of filters in which photometric measurements are available.}
{(5--9)} $B$, $V$, $R$, $I$ and \ha magnitudes and their uncertainties on the last two decimal digits;
{(10)} Total color excess;
{(11)} Effective temperature; 
{(12)} Stellar surface gravity; 
{(13)} Stellar metallicity;
{(14)} Stellar $\alpha$ exponent; 
{(15)} Internal mean photometric offset error;
{(16)} Photometric classification of the source:
(F) Foreground objects,
(RGB) evolved stars consistent with $>100$~Myr PARSEC isochrones,
(MS) Main Sequence stars with no clear \ha  excess,
(Ha) Stars with \ha excess and selected through Eq.~\ref{eq_Ha_selection},
(wHa) Stars associated with \ha$-R < 0$, but not selected through Eq..~(\ref{eq_Ha_selection}) (therefore, weak H-alpha emitters),
(SG) bright objects compatible with blue/red super giants ($B-I > 0.5$ and $V < 16$)
(--) objects associated with incomplete photometry (N$_{filt} < 5$) or that did not fall into any of the previous classification types.}
\end{table*}
\setlength{\tabcolsep}{6pt}

To compare the photometry with the expected brightness of the stars in the SMC, we derived synthetic $\mli{BVRI}$+H$\alpha$ magnitudes of rotating stellar models from O9 to A7-type MS stars. This was achieved by convolving \beatlas photospheric grid of models \citep{beatlas2023}, with the proper transformations to place them on the AB system.

The photospheric \beatlas spectrum is characterized by four parameters: the stellar mass ($M$), the rotation rate $W$ \citep[$v_\mathrm{rot}/v_\mathrm{orb}$,][]{rivinius2013} the MS  lifetime fraction ($t/t_\mathrm{MS}$), and the stellar inclination angle with respect to the line-of-sight ($i$).

To establish the correspondence between each spectral type (ST) to mass (M), we adopted the spectral type versus Mass (ST-M) relation from \citet{schmidt-kaler1982} for B-type stars, with \citet{Adelman2004} and \citet{2005A&A...436.1049M} for A- and O-type stars, respectively.
As an effort to obtain the potential magnitude range for each ST, we set the rotation rate W to its maximum value ($W = 0.99$), as the largest range in brightness occurs when the star is critically rotating and is observed either pole-on ($i=0^{\circ}$) or edge-on ($i=90^{\circ}$).
As for the stellar age, we computed $t/t_\mathrm{MS}$ under the assumption that NGC\,330 has an age of $t=35$~Myr \citep{Bodensteiner20}. For stars earlier than B3, the magnitude range was estimated considering objects at the TAMS ($t/t_\mathrm{MS}=1.0$), as these objects cannot exist in the MS considering the expected age of NGC\,330 and the single stellar evolution scenario.

Given that the \beatlas spectra are dereddened and flux-calibrated for a 10~pc distance, we adopted the values of $E(B-V)=0.11$ mag and $d=57.5$~kpc \citep{milone2018}, with the extinction law as described by \citet{Fitzpatrick1999} with $R_{V}=3.1$ mag.
{Contrary to the relatively unequivocal prediction of the brightness of non-rotating stars, the adoption of rotating models at different inclination angles of the stellar spin axis leads to a relatively wide range of expected magnitudes for a given spectral type, with a non-zero intersection between adjacent subtypes.}
Table~\ref{table_magnitudes_smc} lists the expected magnitude range for $\mli{BVRI}$+\ha filters for the spectral types O9, A0, A2, A7, and all B subtypes. 

\setlength{\tabcolsep}{4pt}
\begin{table}[!ht]
    \begin{center}
    \caption{Expected magnitude range for normal O-, B- and A-type stars in the SMC, including the effects of fast rotation.}
    \label{table_magnitudes_smc}
    \begin{tabular}{c|ccccc}
    \hline
    \hline
    ST & $B$ & $V$ & $R$ & $I$ & H$\alpha$ \\
    \hline
        O9	&   12.9-13.8	&   13.2-14.0		&  13.4-14.2  &  13.8-14.5  &  13.5-14.2     \\
        B0	&  13.4-14.2	&   13.6-14.4		&  13.9-14.6  &  14.2-14.9  &  13.9-14.7    \\
        B1	&  14.7-15.5	&   14.9-15.7		&  15.2-16.0  &  15.5-16.3  &  15.3-16.0   \\
        B2	&  14.9-15.8&   15.2-16.0		&  15.4-16.2  &  15.8-16.5  &  15.5-16.2    \\
        B3	&  15.4-16.2	&   15.7-16.4		&  15.9-16.7  &  16.2-16.9  &  16.0-16.7    \\
        B4	&  16.0-16.8	&   16.2-17.0		&  16.5-17.2  &  16.8-17.5  &  16.5-17.2    \\
        B5	&  16.7-17.5	&   16.9-17.7		&  17.2-17.9  &  17.5-18.2  &  17.2-18.0    \\
        B6	&  17.9-18.7	&   18.1-18.7		&  18.3-19.1  &  18.6-19.3  &  18.4-19.1    \\
        B7	&  18.5-19.4	&   18.7-19.5		&  18.9-19.7  &  19.1-19.9  &  19.0-19.7    \\
        B8	&  18.9-19.7	&   19.0-19.8		&  19.2-20.0  &  19.5-20.2  &  19.3-20.1    \\
        B9	&  19.3-20.2	&   19.4-20.2		&  19.6-20.4  &  19.9-20.6  &  19.7-20.5    \\
        A0	&  19.8-20.7	&   19.8-20.7		&  20.0-20.8  &  20.2-21.0  &  20.1-20.9    \\
        A2	&  20.4-21.4	&   20.4-21.4		&  20.6-21.5  &  20.8-21.6  &  20.7-21.6    \\
        A7	&  20.8-22.0	&   20.8-21.8	&  21.0-21.9  &  21.1-22.0  &  21.1-22.0    \\
    \hline
    \end{tabular}
    \end{center}
\end{table}
\setlength{\tabcolsep}{6pt}

\begin{figure}[!ht]
\includegraphics[viewport=10bp 10bp 520bp 1000bp, clip, width=\columnwidth]{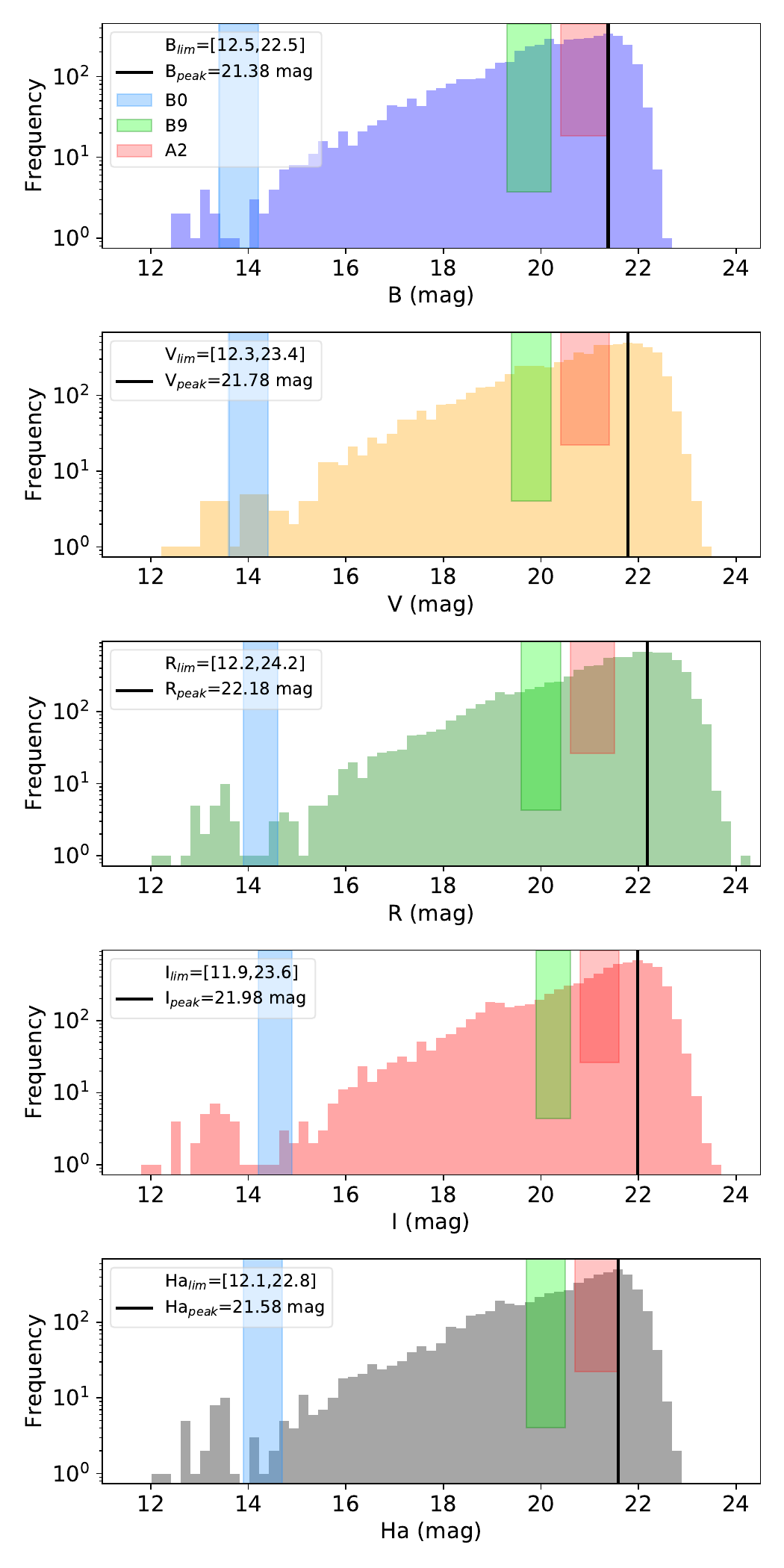}
\caption{Photometric completeness for the $\mli{BVRI}$+{\ha} filters of SAMI observations. Each histogram is sampled with a bin of 0.1~mag, from 11.5 to 24~mag. The vertical dashed line indicates the peak of the distribution. The brightest and faintest object detected on each filter is indicated on the label. We show the expected magnitude intervals for B0 (blue), B9 (green), and A2 stars (orange) for each filter.
\label{fig_photcal_SAMI}}
\end{figure}

The distribution of the calibrated SAMI photometry per filter is presented in Fig.~\ref{fig_photcal_SAMI}. The magnitudes for all filters follow an IMF-like distribution with maximum frequency at 21.3 (B), 21.8 (V), 22.2 (R), 22.0 (I), and 21.6 (\ha), which serves as a proxy for evaluating the photometric completeness limit of the observations.
The completeness of all filters is at least one magnitude fainter than the expected limit for B-type stars (B9, indicated by the green shaded area). The B and {\ha} filters exhibit the shallowest completeness limit, roughly consistent with the expected faintest brightness for A2 stars.
This result highlights that the adopted instrumental setup provides sufficient depth to properly identify all OB and early A-type stars at the distance of the SMC.

\subsection{Comparison with published catalogs}
\label{sect_photometric_comparison}

We validated the outcomes of our photometric calibration, as detailed in Sect.~\ref{sec_cal_templatefitting}, by cross-referencing the results with three previously published catalogs: the Optical Gravitational Lensing Experiment \citep[OGLE-II,][]{Udalski98}, the Magellanic Clouds Photometric Survey \citep[MCPS,][]{Zaritsky04}, and the AO-assisted MUSE observations from \citet{Bodensteiner20}.
For the cross-matching procedure, we adopted a maximum separation of 0\farcs5 for the seeing-limited OGLE and MCPS positions and 0\farcs3 for the AO-assisted MUSE positions, leading to a sample of 3230 (OGLE), 1341 (MCPS), and 250 (MUSE) sources with SAMI counterparts.

\begin{figure}[t!]
\centering
\includegraphics[viewport=15bp 15bp 550bp 400bp, clip, width=\columnwidth]{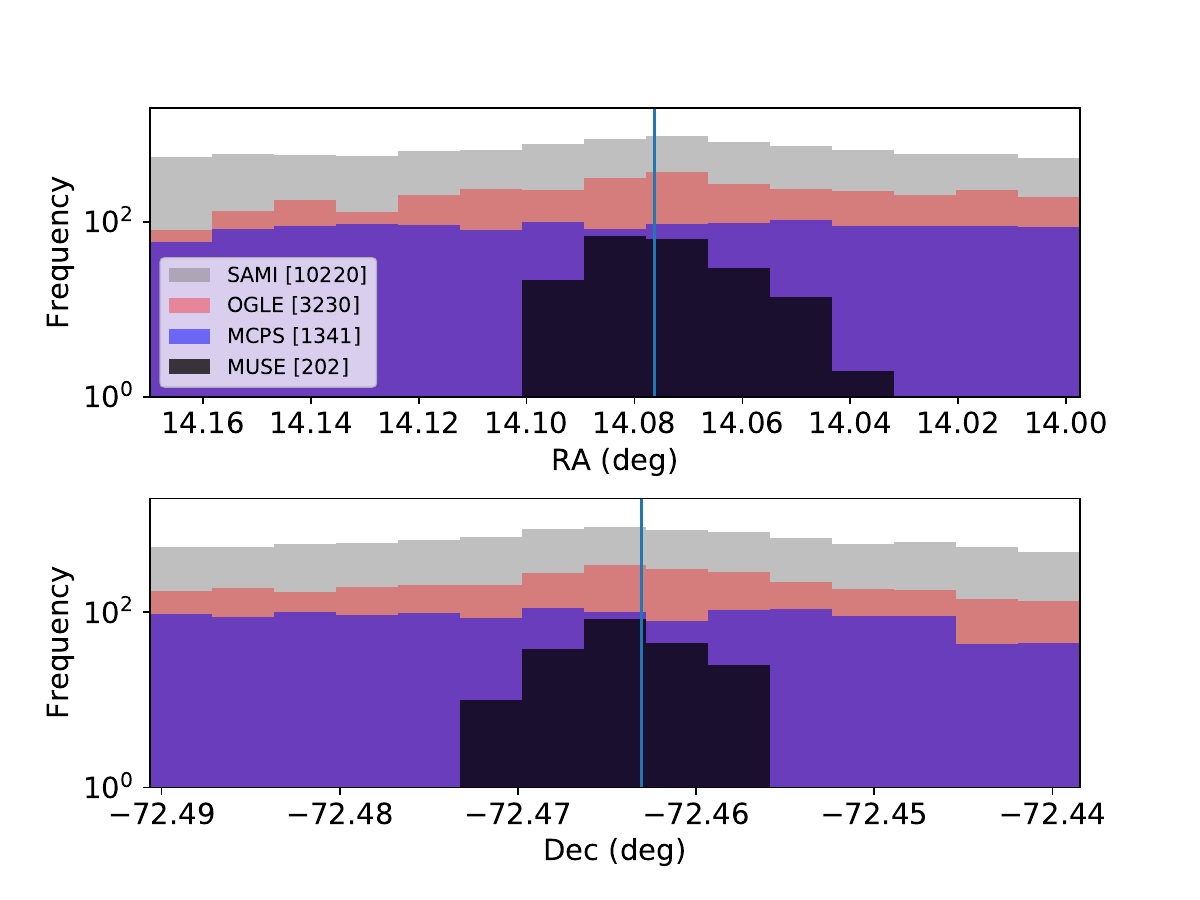}\\[-1.0ex]
\caption{Distribution of sources in the NGC\,330 FOV from the SAMI catalog (grey), and the external catalogs (red: OGLE, blue: MCPS, black: MUSE) in the RA (top panel) and declination directions (bottom). The number of sources from each catalog is indicated on the label.
The vertical blue line indicates the central coordinates of NGC\,330.
\label{fig_pos_ngc330_extcat}}
\end{figure}

Figure~\ref{fig_pos_ngc330_extcat} presents the spatial distribution of the sources of the SAMI catalog and the published catalogs (red: OGLE; blue: MCPS; black: MUSE) along the RA  (top panel) and declination (bottom) directions.
Far from the cluster's center, OGLE and MCPS exhibit a relatively constant distribution of sources in both RA and DEC directions. An overdensity of OGLE sources is observed towards the cluster's central position, which is not evident for the MCPS, indicating that OGLE has a better completeness towards NGC\,330 than MCPS.
MUSE observations are limited for $V \leq 18.5$~mag objects, located within a considerably smaller {$\sim$1\arcmin$\times$1\arcmin} FOV around the cluster's center. Consequently, MUSE covers a limited area, capturing a relatively smaller number of targets than the other catalogs.

\begin{figure}[ht]
\centering
\includegraphics[width=\columnwidth]{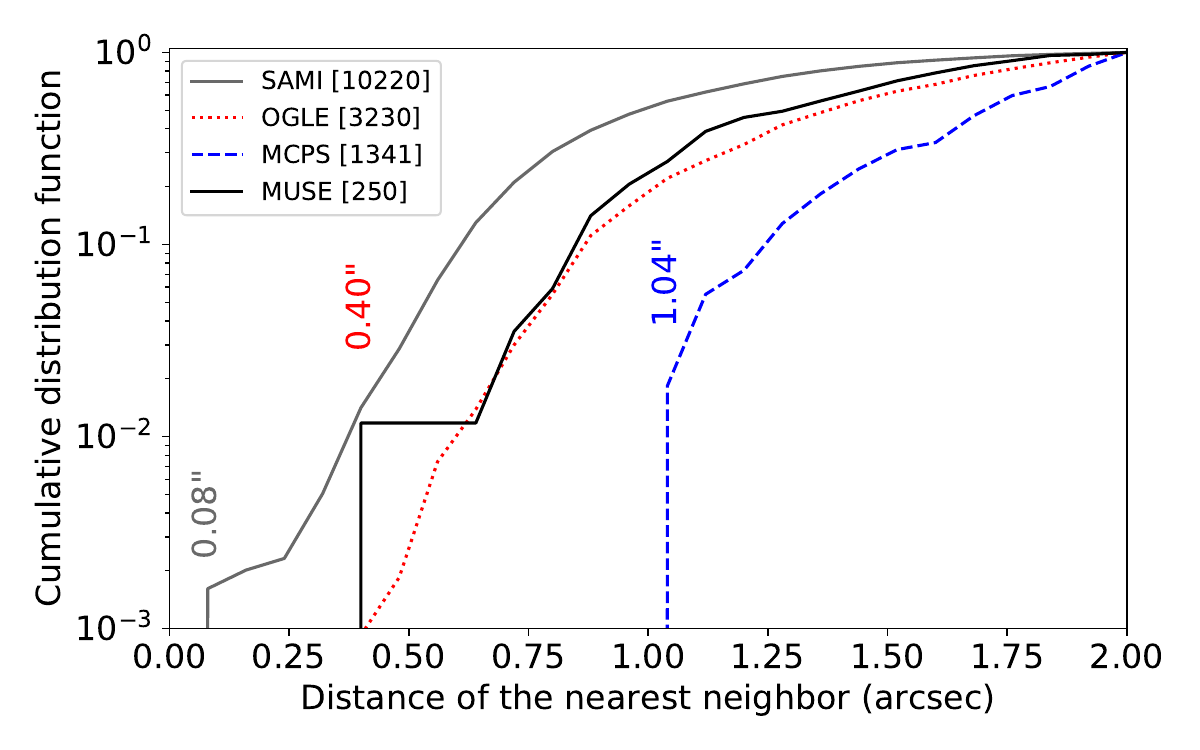}\\[-2.0ex]
\caption{
Comparison of the distance of the nearest neighbor from different photometric catalogs: SAMI (filled grey curve), OGLE (dotted red), MCPS (dashed blue), MUSE (filled blue).
The curves represent the cumulative distribution function of the quantity, starting at zero and progressing towards one. The total number of objects in each respective catalog is indicated on the plot.
The minimum separation between the sources of each catalog is indicated in the plot. 
\label{fig_minsep}}
\end{figure}

Figure~\ref{fig_minsep} displays the cumulative distribution function of the distance of the nearest neighbor in each of the four catalogs.
This metric represents the likelihood of a single object in one catalog being matched with one or multiple sources from another catalog{. It is particularly significant} when comparing catalogs derived from seeing-limited observations (such as OGLE and MCPS) with those from adaptive optics-assisted observations (like MUSE and SAMI), particularly in crowded areas like the central region of the cluster.
This quantity also relates to the depth of the images, as fainter targets detected in one catalog might be absent in another.
Consequently, the probability of cross-matching multiple objects from a higher-resolution catalog (e.g., SAMI) with a single object from a catalog with a coarser resolution increases as this distance gets larger.
For instance, a {single bright} object in the seeing-limited OGLE or MCPS catalogs can be resolved into multiple{, fainter} objects detected in SAMI images.
The minimum distance corresponds to 1\farcs04 for MCPS, and 0\farcs40 for OGLE and MUSE. A minimum separation of 0\farcs08 is observed for SAMI, indicating a significant gain in resolution and depth from the AO-assisted observations.

We now compare the calibrated $\mli{BVI}$ magnitudes with those from the OGLE catalog to check the reliability of our photometric calibration.
OGLE magnitudes in the Vega system were converted to the AB system, and comparisons were made for sources brighter than 20~mag in all the $\mli{BVI}$ filters.
The OGLE catalog was primarily selected given that, among the three published catalogs, OGLE exhibits the deepest completeness of sources in the same FOV of SAMI observations (Fig.~\ref{fig_pos_ngc330_extcat}) and displays a relatively small minimum separation of closest neighbors (0\farcs4, Fig.~\ref{fig_minsep}).
For completeness, the comparison between SAMI photometry and the MCPS and MUSE catalogs is presented in Appendix~\ref{appendix_catalogs}.

\begin{figure}[!t]
\centering
\includegraphics[width=\columnwidth]{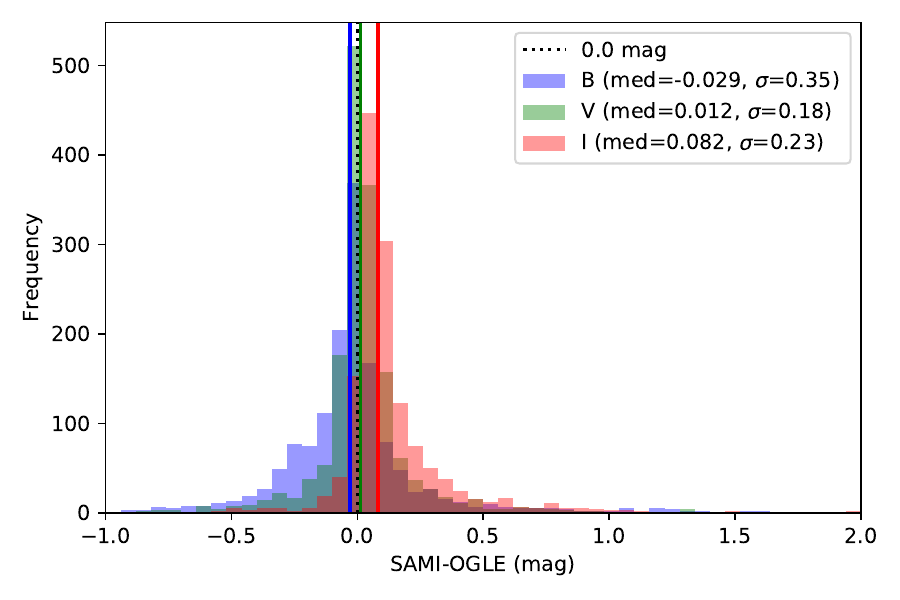} \\[-2.5ex]
\caption{Distribution of residuals between $\mli{BVI}$ magnitudes from SAMI and OGLE catalogs. Filled vertical lines indicate the median values of the distributions, while the dashed line is placed at $x=0$. The median value and 1-$\sigma$ width (68\%) of each distribution are indicated in the label.
\label{fig_photcal_residuals_OGLE}}
\end{figure}

The distribution of the residuals between SAMI and OGLE magnitudes are presented in Fig.~\ref{fig_photcal_residuals_OGLE}. The residuals are distributed around 0~mag with 1-$\sigma$ widths smaller than 0.4~mag in all filters.
Despite the relatively large widths, the comparison with OGLE gives the most symmetric results among the three external catalogs.
Two factors can explain the extended tails in the distributions:
$i)$ deblending of multiple sources, as we are comparing  photometry from AO-assisted observations with coarser seeing-limited catalogs; and 
$ii)$ intrinsic variability among sources in the sample (e.g., CBes), given the significant time gaps between the observations.

The results from this section and Appendix~\ref{appendix_catalogs} indicate that the adopted calibration method is robust and consistent with the published catalogs, providing calibrated AB magnitudes, ranging between $\sim$12 and $\sim$24~mag, for all $\mli{BVRI}+$H$\alpha$ filters.

\subsection{Identification of foreground sources from Gaia DR3}
\label{sect_gaia_foreground}

We took advantage of the astrometric stellar parameters from Gaia Data Release 3 \citep[Gaia\,DR3,][]{gaiadr3} to identify and remove Galactic foreground sources in the SAMI FOV. Initially, we cross-matched the SAMI catalog with Gaia\,DR3, leading to a list of 2866 sources with Gaia counterparts.
Given our photometric completeness extending beyond Gaia\,DR3, the faintest objects associated with a Gaia\,DR3 counterpart {exhibit} magnitudes reaching $V \lesssim 22.0$~mag, about $\sim1.5$~mag brighter than the faintest objects detected by SAMI (cf. Table~\ref{table_psf_fitting} and Fig.~\ref{fig_photcal_SAMI}).

Subsequently, we implemented the following selection criteria:
\begin{itemize}
\item Removal of sources associated with large {Renormalized Unit Weight Error} values, $\mathrm{RUWE} > 1.4$ \citep{ruwe2021}, keeping only those associated with a good astrometric solution;
\item Removal of sources exhibiting parallaxes with nominal values smaller than 1-$\sigma$; and 
\item Selection of sources exhibiting parallax values $|\varpi| \geq 0.1$~mas, much larger than the expected parallax value for the SMC, $\omega_\mathrm{SMC} \approx 0.016$~mas, estimated from the distance of $d_\mathrm{SMC} = 62.44$~kpc \citep{distancesmc}.
\end{itemize}
Employing the above procedure resulted in 849 confirmed foreground objects ($\approx 8\%$ of the SAMI catalog), which were excluded from the results and analysis presented in the next section.

\section{Results}
\label{sec_results}

\subsection{Photometric analysis of \texorpdfstring{NGC\,330}{NGC 330}}
\label{sec_results_sami_ngc330}

\begin{figure*}[!ht]
\centering
\includegraphics[viewport=0bp 0bp 800bp 400bp, clip, width=0.9\textwidth]{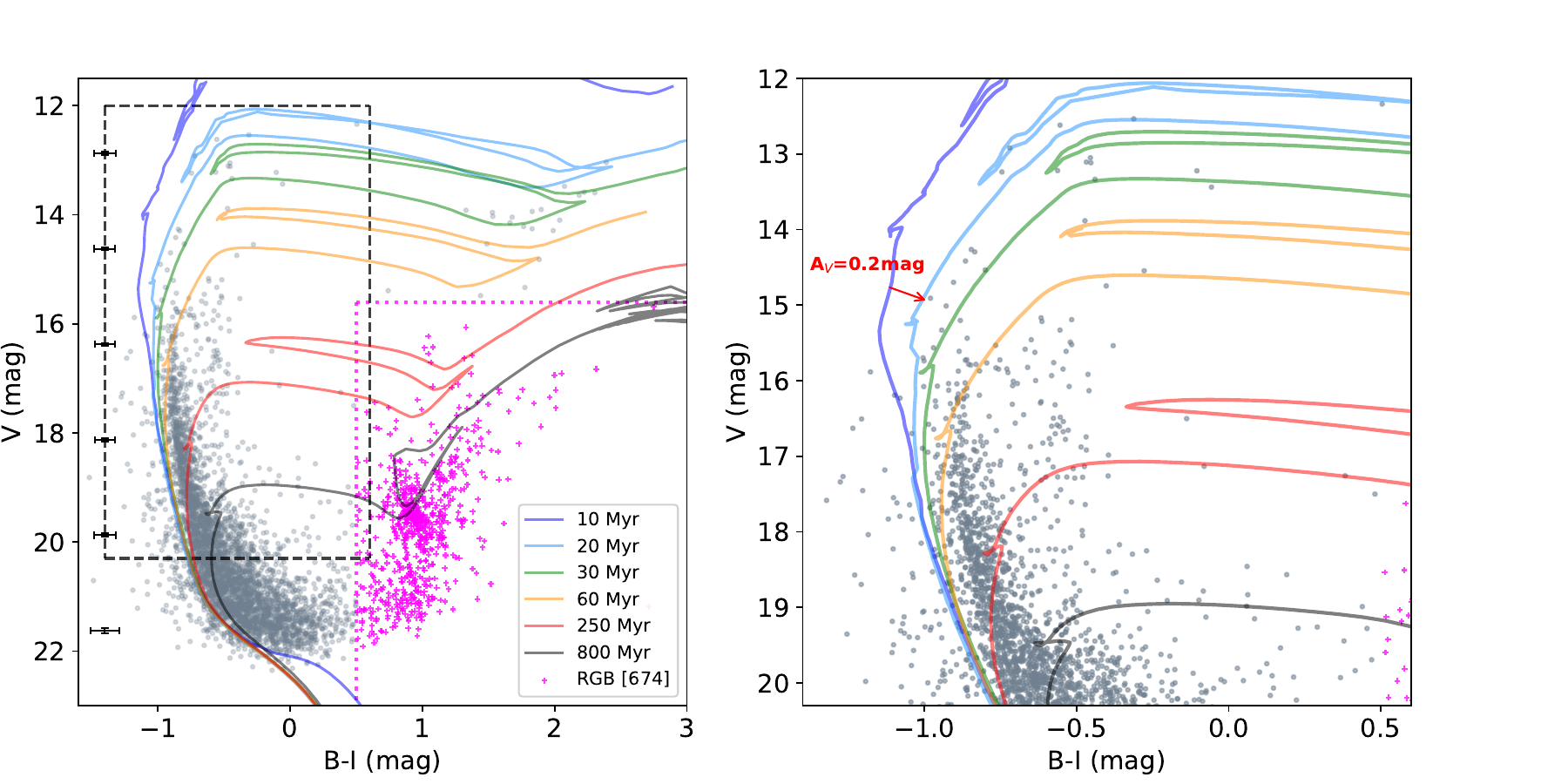}\\[-1.0ex]
\caption{{Color-magnitude diagram based on the SAMI photometry for NGC\,330.
PARSEC evolutionary tracks for ages between 10 and 800~Myr are overlaid in different colors.
{Magenta $+$ symbols within the dotted lines} are likely associated with an old stellar population, consistent with the 250-800~Myr isochrones. 
The black error bars indicate the typical errors of the points as a function of the $y$-axis.
The right panel shows a zoom-in view of the main sequence{, delimited by the black dashed lines in the left panel}.
The red arrow indicates the direction of the reddening vector, assuming $A_{V}=0.2$~mag.}
\label{fig_cmd_ccd_ngc330_isochrone}}
\end{figure*}

We used the calibrated $\mli{BVI}$ photometry, excluding foreground objects, to construct the C-M diagram presented in Fig.~\ref{fig_cmd_ccd_ngc330_isochrone}.
PARSEC \citep{parsec} evolutionary tracks for ages ranging from 10 to 800~Myr are overlaid with distinct colors to assist the analysis. The isochrones were scaled to the distance of the SMC ($m-M = 18.97$~mag, \citealt{milone2018}) and a $A_{V}=0.2$~mag reddening vector, assuming the \citet{cardelli1989} extinction law with $R_{V}=3.1$, is overlaid.

Most sources are located on the main sequence at $B-I \lesssim 0$~mag, while a considerable fraction of targets exhibit redder colors ($B-I > 0.3$~mag). 
Sources brighter than $V = 16 $~mag are consistent with a population of main sequence (for $B-V < 0$~mag) and evolved objects ($B-V \gtrsim 0$~mag) with ages between 20 and 60~Myr.
The right panel shows that the top-brightest sequence of stars is consistent with the 10~Myr isochrone (dark blue), assuming $A_{V}=0.2$~mag (indicated by the reddening vector).
Targets at $V \sim 15$~mag and $B-I \sim 2$~mag are consistent with a relatively older population with ages between 50 and 60~Myr (yellow isochrone).
We also observe a relatively large fraction of stars located to the right of the main sequence (see details on the right panel), which are likely objects associated with {\ha} excess at the expected position of CBe stars.

Fainter targets exhibiting larger $B-I$ values represent an older population of evolved stars in the Red Giant Branch (RGB) phase and, therefore, are clearly not associated with the young cluster.
We identified 674 sources fainter than $V=15.5$~mag and exhibiting $B-I > 0.5$~mag (shown as {magenta $+$ symbols}). These sources were not filtered out based on their Gaia DR3 parameters, suggesting they are either foreground Galactic objects or field stars from the SMC with poor astrometry (see Sect.~\ref{sect_gaia_foreground}) and will therefore be omitted from further analysis. 
A mixture between young main sequence and older stars leaving the main sequence is likely observed for $V \gtrsim 18$~mag, making it difficult to disentangle different populations solely based on color-criteria selections.

\begin{figure*}[ht!]
\centering
\includegraphics[viewport=0bp 0bp 930bp 400bp, clip, width=\textwidth]{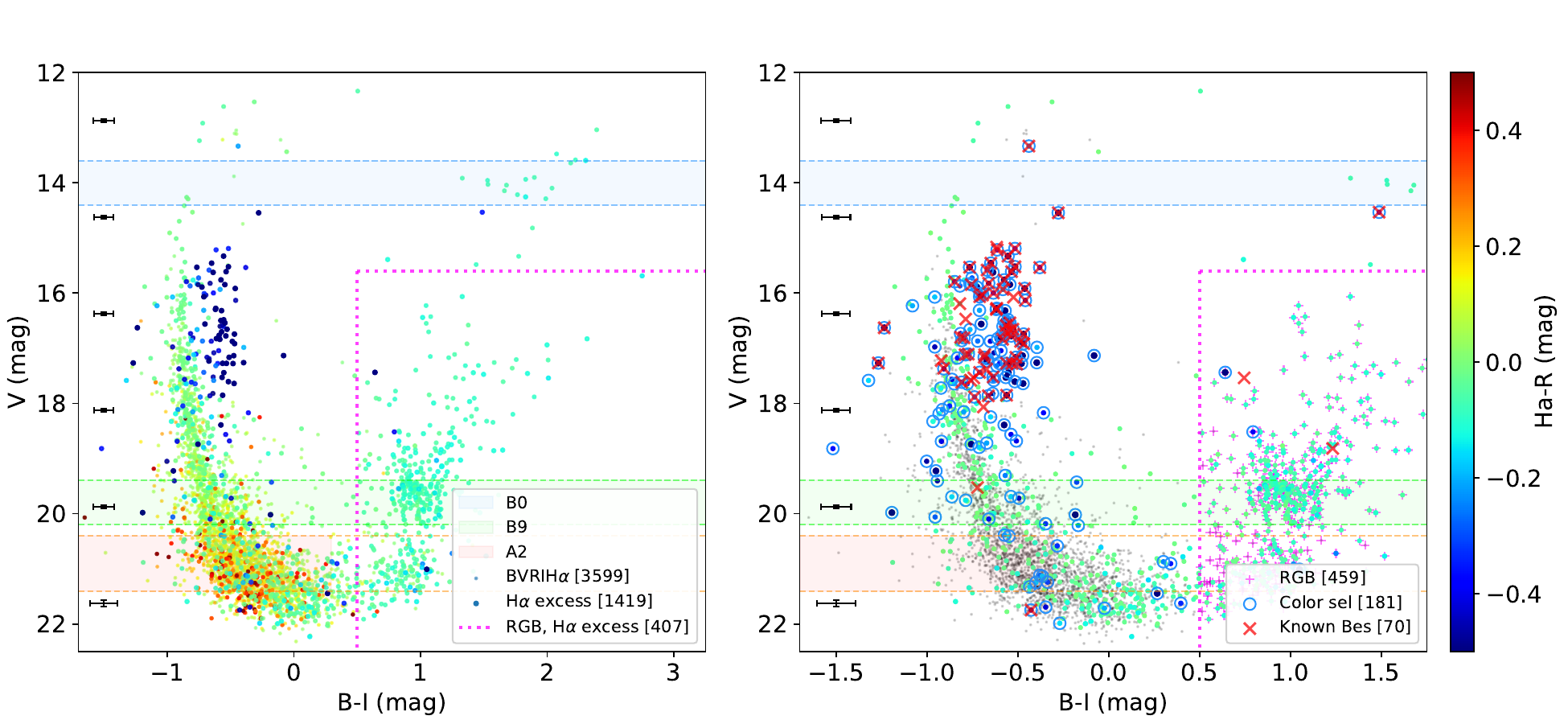}\\[-1.0ex]
\caption{(left) $V$ vs $B-I$ color-magnitude diagram of NGC\,330 with SAMI photometry. The points are color-coded using the H$\alpha-R$ color index. The total number of sources with H$\alpha$ excess (H$\alpha-R < 0$) is indicated on the label. The number of old stars (located within the {magenta dotted lines}) with H$\alpha$ excess is also indicated.
The expected magnitude intervals for B0 (blue), B9.5 (green), and A2 stars (red) are represented by the shaded regions.
The black error bars on the left indicate the typical errors of the points as a function of the $y$-axis.
(right) An inset of the color-magnitude diagram highlighting the sources exhibiting H$\alpha$ excess. The points associated with H$\alpha-R < 0$ are color-coded, while points with H$\alpha-R > 0$ are indicated as smaller {grey} dots. 
Red $\times$ symbols correspond to known CBe stars from previous works, and blue circles indicate the H$\alpha$ emitters selected through the color selection criterion (see text). {Magenta} $+$ symbols indicate stars in the RGB phase (unlikely members of NGC\,330).
\label{fig_cmd_ngc330}}
\end{figure*}

The same C-M diagram is presented in Fig.~\ref{fig_cmd_ngc330} (left), with each point color-coded based on its H$\alpha-R$ color index, taking into advantage the calibrated {\ha} and $R$ photometry.
This color index is a variation of the $R-$\ha index, often adopted as a proxy for selecting \ha emitters in the SMC \citep[e.g.,][]{Keller99, Iqbal13}, which may include stars undergoing the Be phenomenon, Red Supergiants (RSGs), Blue Supergiants (BSGs), and possibly active young stellar objects (YSOs).
We further compare the expected $V$-magnitude range for B0, B9, and A2 stars listed in Table~\ref{table_magnitudes_smc} as shaded regions over the diagram.

A total of 1419 sources exhibiting {\ha} excess (that is, H$\alpha-R<0$~mag) are shown as green to blue points in the C-M diagram. The visual inspection of the C-M diagram suggests four main groups of {\ha} emitters:
\begin{enumerate}
    \item The bulk of {\ha} emitters (blue points) are located to the right of the main sequence at $B-I<0$~mag, spread around the expected $V$-band magnitudes for B-type stars (indicated by the blue and green regions) and consistent with the expected loci of CBes (e.g., \citealt{Keller99} and \citealt{Iqbal13});
    \item Several {\ha} emitters are detected to the left of the main sequence, aligning with the expected position of blue stragglers or mergers (e.g., \citealt{Dresbach22}); and
    \item A third group corresponds to red and evolved objects stars ($V > 15.5$, $B-I>0.6$) located within the RGB region defined in Fig.~\ref{fig_cmd_ccd_ngc330_isochrone}, for which 407 of 459 objects ($88.7\pm4.4\%$) exhibit H$\alpha-R < 0$. The behavior for more evolved objects to exhibit a relatively large {\ha} excess compared to typical MS stars has been previously observed in earlier studies of NGC\,330 (e.g., see Fig.~1 from \citealt{Keller99} and Fig.~2 from \citealt{Iqbal13}) and other clusters (e.g., NGC\,6752, \citealt{Bailyn96}).

    \item We note a significant number of sources with a positive H$\alpha-R$ color in the faintest region of the MS ($V \gtrsim 20$~mag). Considering the shallower photometric completeness of the {\ha} filter ($\sim21.7$~mag, Fig.~\ref{fig_photcal_SAMI}) in comparison to the $R$ filter ($\sim 22.2$~mag), it is possible that these positive H$\alpha-R$ values could be a result of a signal-to-noise effect.
\end{enumerate}

\subsection{Selection of \texorpdfstring{\ha}{H-alpha} emitters}

Previous studies \citep[e.g.,][]{Keller99,Iqbal13} adopted $R-$H$\alpha$ versus $V-I$ diagrams to identify active CBes in NGC\,330. By adopting a simple color selection criteria, they were able to separate objects exhibiting clear  {\ha} excess from normal stars, thus excluding the bulk of main sequence stars and the cooler evolved objects.
Although both studies have made significant contributions to identifying CBe stars in the SMC, their findings were limited by the following observational aspects:
\begin{enumerate}
    \item Their photometric analysis was based on seeing-limited observations (1\farcs5--2\farcs0), not sufficient to disentangle and resolve the stellar population in the innermost and crowded central region of the cluster;
    \item Both works limited their sample to relatively bright stars ($V < 17.5$~mag for \citealt{Keller99} and $V < 18.0$~mag for \citealt{Iqbal13});
    \item Arbitrary zero-points were adopted for the $R-$H$\alpha$ color index (so that normal MS stars had $R-${\ha}$\approx0$).
\end{enumerate}

{HST observations presented by \citet{milone2018} were able to overcome the coarser angular resolution of previous observations from \citet{Keller99,Iqbal13}; however, their analysis did not show significant {\ha} emission detected for $I$-band magnitude around $\sim 18$~mag (see their Fig.\,2), corresponding to B6 stars at the distance of the SMC.

Leveraging the superior spatial resolution provided by the adaptive optics AO-assisted SAMI observations {with sufficient photometric depth in all broad-band $\mli{BVRI}$ and narrow-band {\ha} filters}, we have reevaluated the color selection criteria utilized previously {by \citet{Keller99} and \citet{Iqbal13}}. This reassessment was undertaken to validate our ability to identify both known and new {\ha} emitters ({most probably} active CBes) within NGC\,330.

We first cross-matched the positions of known CBe stars with the SAMI catalog, leading to a total of 70 unique objects, from which 47, 31 and 23 are from \citet[][K99]{Keller99}, \citet[][I18]{Iqbal13}, and \citet[][M07]{martayan2007}, respectively (12 objects are in common with K99 and I13, while four objects are unique from M07). They are indicated as red $\times$ symbols in Figs.~\ref{fig_cmd_ngc330} (right) and \ref{fig_ccd_ngc330_Ha_sel}.

\begin{figure}[!ht]
\includegraphics[viewport=0bp 10bp 510bp 440bp, clip, width=\columnwidth]{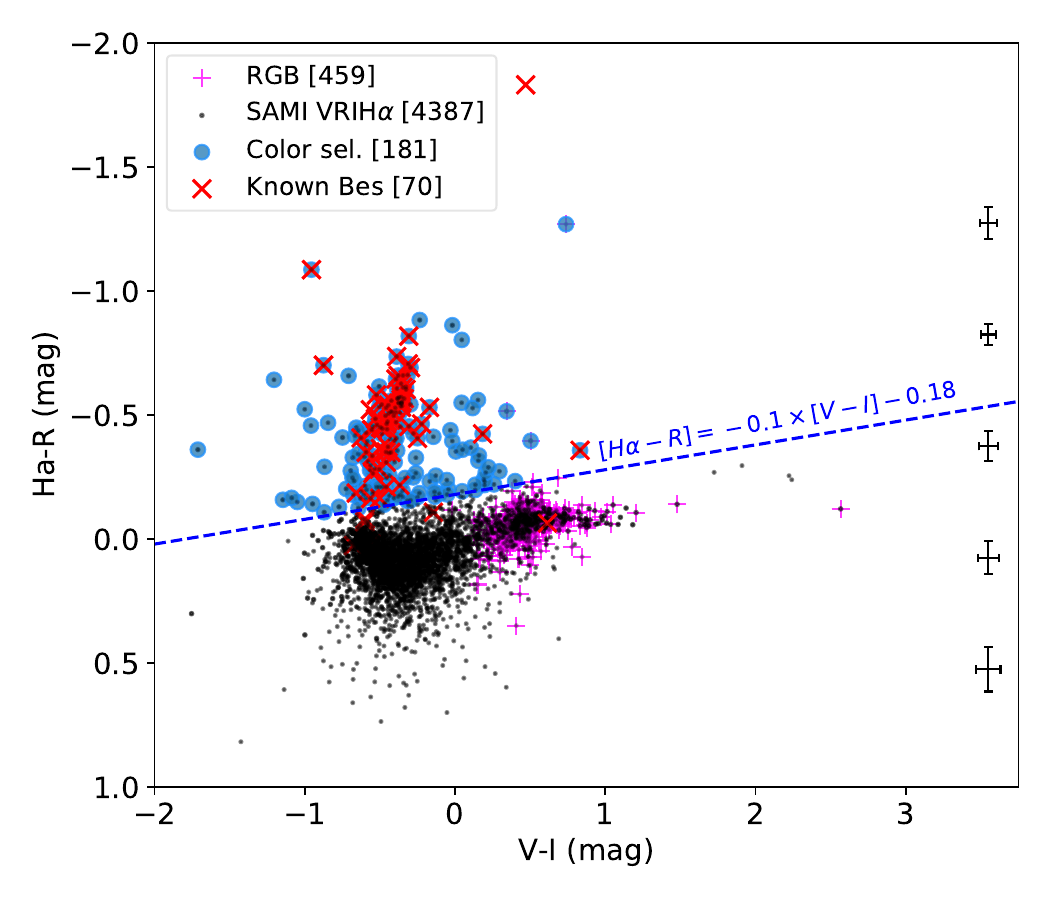}
\caption{H$\alpha-R$ vs. $V-I$ color-color diagram for NGC\,330. Red $\times$ symbols correspond to known CBe stars from previous works (see text). Blue circles indicate the H$\alpha$ emitters selected through the color selection criterion defined by {Eq.~(\ref{eq_Ha_selection}) and indicated by the dashed blue line}. {Magenta} $+$ symbols {are likely} stars in the RGB phase (not associated with NGC\,330). The black error bars indicate the typical errors of the points as a function of the H$\alpha-R$ color.}
\label{fig_ccd_ngc330_Ha_sel}
\end{figure}

The overall distribution of the points in Fig.~\ref{fig_ccd_ngc330_Ha_sel} exhibits a clear concentration of targets with positive H$\alpha-R$ values, consistent with main sequence stars with $V-I \sim 0$~mag) and evolved red objects at $V-I \gtrsim 0.5$~mag (which is also consistent with the position of most of the sources classified as RGB sources). 
The sources exhibiting {\ha} excess, however, have $V-I$ colors ranging from -2.0 (to the left of the MS) to 1.0~mag (to the right of the MS).
As expected, almost all known CBe stars (red $\times$ symbols) are located at the top of the C-C diagram, exhibiting negative H$\alpha-R$ values. Only three targets exhibit H$\alpha-R \approx 0$~mag values: two are located within the MS objects, and one is found within the RGB sequence at $V-I \approx 0.5$~mag. 

To expand the selection of similar objects exhibiting {\ha} excess, we adopted the following color selection criterion:
\begin{equation}
    (\mathrm{H}\alpha-R) = -0.10 \cdot (V-I) - 0.18\,,
    \label{eq_Ha_selection}
\end{equation}
\noindent indicated as the {dashed blue} line in Fig.~\ref{fig_ccd_ngc330_Ha_sel}, such that objects with (H$\alpha-R$) smaller than the above criterion for a given $(V-I)$ are considered to have {\ha} excess.
This color selection criterion identified 181 sources exhibiting {\ha} excess, represented by the blue circles in the C-C diagram from Fig.~\ref{fig_ccd_ngc330_Ha_sel}.
This criterion roughly lies between the color selection criteria established by \citet{Keller99}\footnote{H$\alpha-R \approx -0.05 \cdot (V-I)-0.33$, in AB magnitudes} and \citet{Iqbal13}\footnote{H$\alpha-R \approx -0.15 \cdot (V-I)-0.04$, in AB magnitudes}. Moreover, it also considers the higher quality and the deeper photometric depth achieved by SAMI observations compared to the previous works.

Naturally, the adoption of Eq.~(\ref{eq_Ha_selection}) led to a much smaller number of objects than by just selecting all sources exhibiting negative H$\alpha-R$ values (1419, Fig.~\ref{fig_cmd_ngc330}). In addition, it correctly identifies most of the known CBe stars (67 of 70),  excluding the majority of evolved objects to the right of the MS (indicated as green $+$ symbols).
The three known CBe stars not selected by the color criterion of Eq.~(\ref{eq_Ha_selection}) are located within the MS locus. This is expected, given the long-term variability of CBe stars. In a time scale of a decade, some stars may have dissipated their disks (entirely or partially) with a corresponding reduction of their {\ha} excess.

The right panel of Fig.~\ref{fig_cmd_ngc330} highlights the sources associated with negative \ha$-R$ indices, and also shows the known CBe stars and the objects selected through the color criterion from Eq.~(\ref{eq_Ha_selection}).
The known CBe stars are relatively bright objects ($V \lesssim 18$~mag), given the flux-limited nature of the previous observations. The sources selected using Eq.~(\ref{eq_Ha_selection}) are observed over the full range of $V$-band magnitudes (i.e., luminosities).
A total of 137 {\ha} emitters selected through the color criteria exhibit $V < 20.3$~mag, corresponding to stars earlier or equal to B9. The remaining 44 sources exhibit $20.3 < V \lesssim 22.5$~mag, as expected for lower mass objects with spectral types ranging from early to mid-A stars.

The exact nature of these low-mass stars associated with {\ha} excess is difficult to assess. At least for some of them, the {\ha} excess is spurious due to the larger photometric errors in the {\ha} filter. However, eight sources exhibit positive $B-I$ indices, consistent with objects leaving the MS through the red-giant branch (RGB).

According to the color selection criterion, most of the bright {\ha} emitters ($V \leq 20.3$~mag) are located to the right of the MS, together with most of the known CBe stars from previous works.
About 16 {\ha} emitters (including two known CBe stars) are identified to the left of the MS ($B - I \lesssim -1$~mag), corresponding to stellar mergers or blue stragglers candidates.
Only four {\ha} emitters are associated with $B - I$ values larger than 0.5~mag, three are likely old RGB sources, and the brightest one ($V=14.535 \pm 0.026$~mag, {\#45 in Table\,\ref{table_phot_SAMI}}) corresponds to a known binary system {associated with a cold supergiant component} \citep[{SMC5\_002807,}][]{martayan2007}.

{We compared the total number of 63 {\ha} emitters identified by \citet[][see their Fig.\,2]{milone2018} showing F814W\,$\lesssim$\,18\,mag with the equivalent number of 83 {\ha} emitters from our selection exhibiting $I$\,$<$\,18\,mag. The difference of 20 sources detected in SAMI observations, but not in HST data, can be attributed to two main factors:
$i)$ the shallower photometric completeness of HST observations; and/or
$ii)$ the intrinsic variability of the Be phenomenon over a 1.5-year scale. HST observations date back to July 2017 (see their Table 1), while SAMI observations are from January 2019 (Sect.~\ref{sec_obs}).}

\subsection{Comparison with MUSE}
\label{sect_muse}

The recent MUSE observations of the central region of NGC\,330 ($\alpha = 00$h\,$56$m\,$16.8$s, $\delta = -72^{\circ}\,27\arcmin\,47\arcsec$, FOV$\sim$63\arcsec$\times$63\arcsec) reported by \citet{Bodensteiner20} offers an excellent opportunity to compare spectroscopic and photometric results.
{The identification of Be stars in the study by \citet{Bodensteiner20} was based on the analysis of the MUSE spectra, while the $V$-band magnitudes adopted in their work were extracted from both ground-based and HST photometric catalog.}

First, we selected all the 1242 objects from the SAMI catalog that lie within the MUSE FOV (from here on, it will be referred to as the ``SAMI subsample''). Fig.~\ref{fig_completeness_muse} presents the photometric completeness of this subsample, confirming that it is also complete within the expected magnitude ranges for B-type stars in all filters. For instance, the photometric completeness for the $V$ band is at $V = 20.94$~mag, about $\sim 2.5$~mag deeper than the $V = 18.5 $~mag limit reached by \citet{Bodensteiner20}. A total of 334 sources with SAMI photometry are brighter than $V = 18.5$~mag (referred to as the ``bright subsample''), while 908 are fainter than the photometric limit adopted throughout their analysis.

\begin{figure}[!ht]
\includegraphics[viewport=10bp 10bp 510bp 1000bp, clip, width=\linewidth]{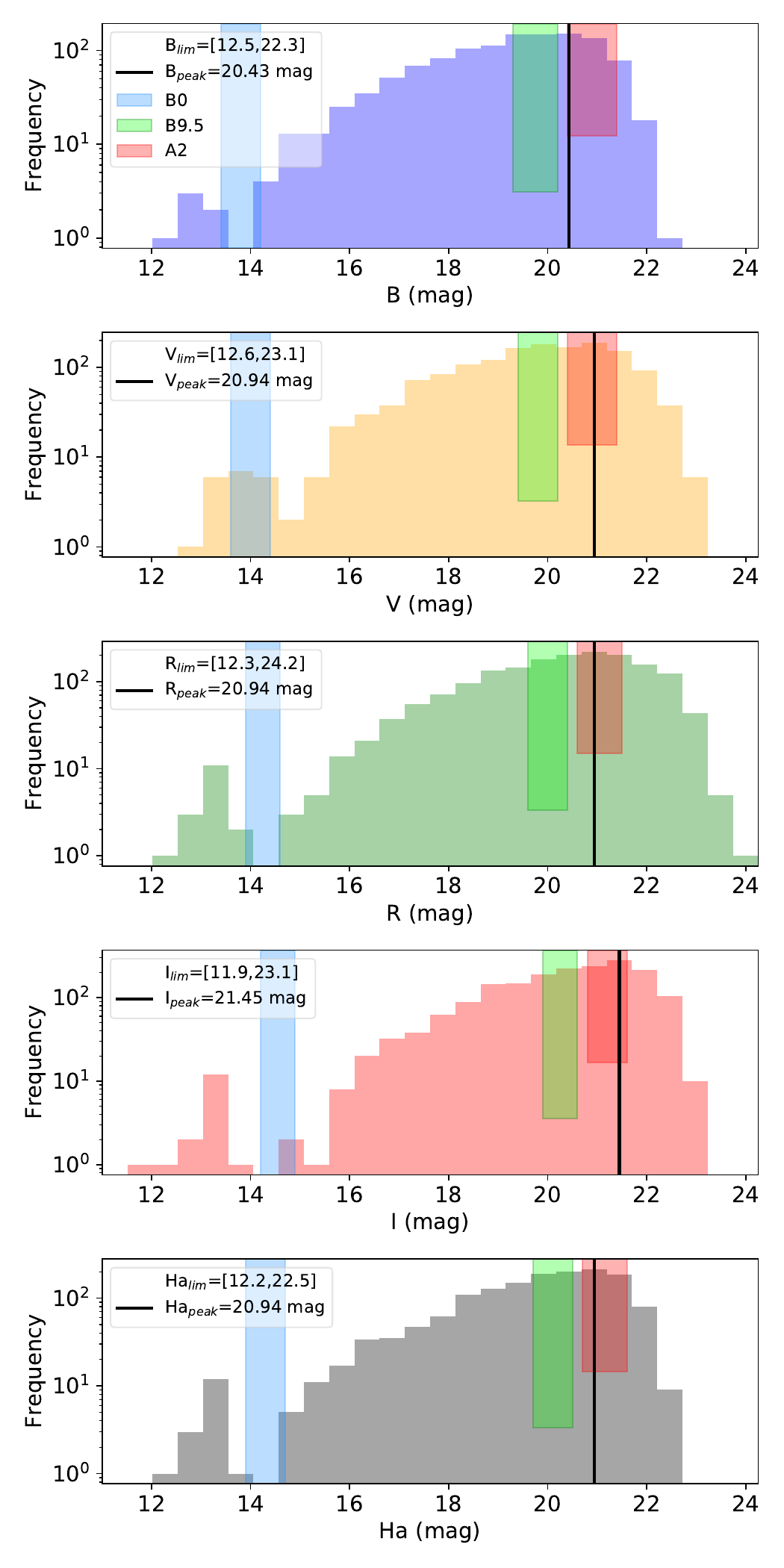}
\caption{Same as Fig.~\ref{fig_photcal_SAMI} but for a smaller FOV of {1\arcmin$\times$1\arcmin} centered at $\alpha = 00$h\,$56$m\,$16.9$s, $\delta = -72^{\circ} 27$\arcmin$48.7$\arcsec, and bins of
0.4~mag comprising the same magnitude range.}
\label{fig_completeness_muse}
\end{figure}

A cross-match between the SAMI catalog {and} the objects listed in Table~D.1 of \citet{Bodensteiner20} resulted in 250 targets with measurements in all five filters (henceforth ``MUSE subsample''). From these, 73, 163, and 14 were classified by \citeauthor{Bodensteiner20} as Be, normal B, and giants (RSG, BSG), respectively.

To identify the {\ha} emitters within the MUSE FOV, we applied the same selection criterion from Fig.~\ref{fig_ccd_ngc330_Ha_sel} on the SAMI (1242 sources) and the MUSE subsamples (250 sources). The results are presented in Fig.~\ref{fig_ccd_ngc330_Ha_sel_museFOV}.

\begin{figure}[!ht]
\includegraphics[viewport=10bp 10bp 510bp 860bp, clip, width=\linewidth]{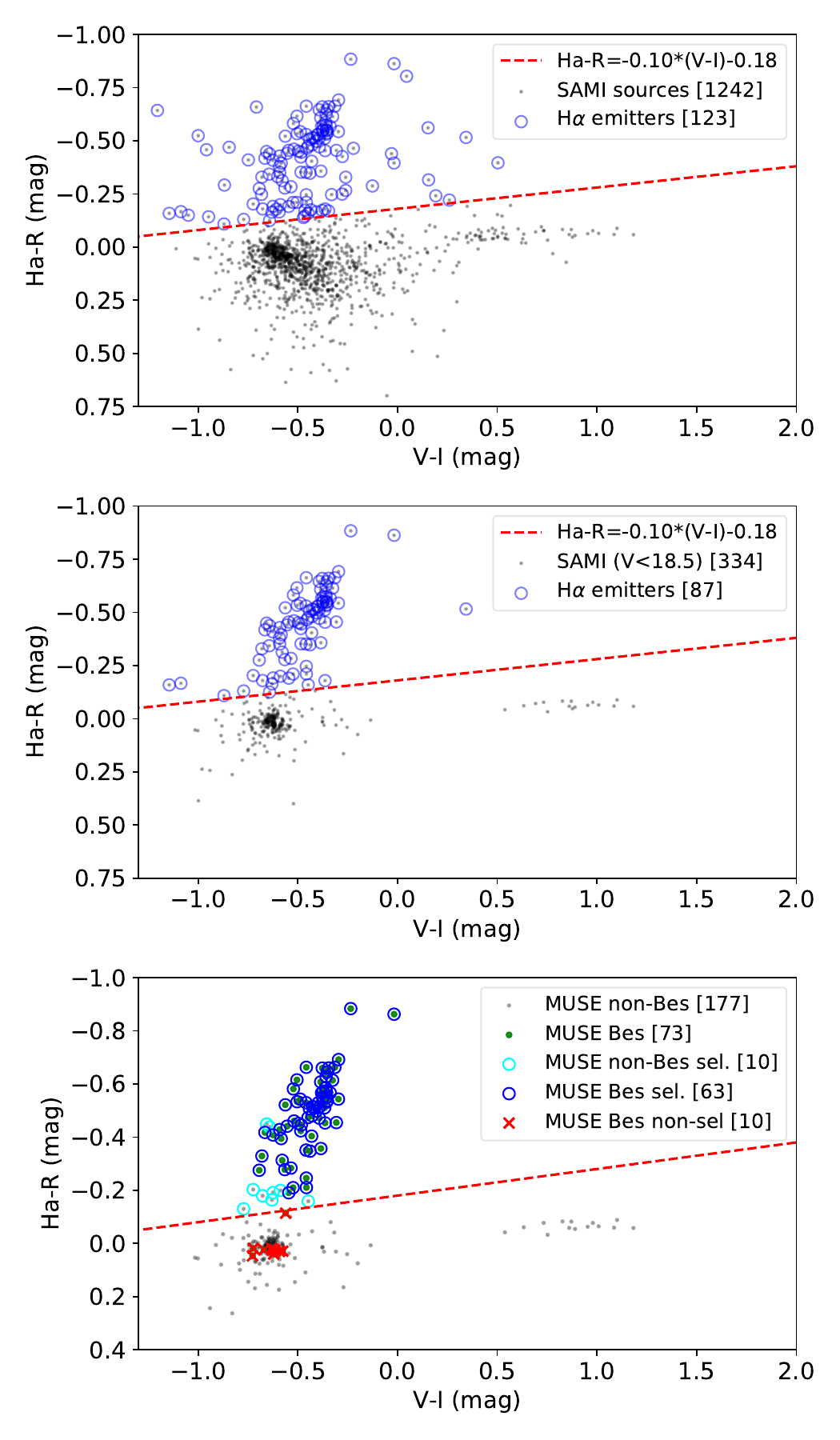}
\caption{Same as Fig.~\ref{fig_ccd_ngc330_Ha_sel} but for a smaller FOV of {63\arcsec$\times$63\arcsec} centered at $\alpha = 00$h\,$56$m\,$16.9$s, $\delta = -72^{\circ} 27$\arcmin$48.7$\arcsec.
Top panel: 1242 sources with SAMI photometry.
Middle: flux-limited ($V \leq 18.5$~mag) subsample of the SAMI sources.
Bottom: 250 MUSE objects detected on the SAMI catalog.
Open blue circles correspond to the {\ha} emitters selected through the color selection criterion of Eq.~(\ref{eq_Ha_selection}), shown as dashed red lines.
Open cyan circles correspond to non-Be MUSE objects exhibiting {\ha} excess. Red $\times$ symbols indicate spectroscopic confirmed CBe stars not showing {\ha} excess.
\label{fig_ccd_ngc330_Ha_sel_museFOV}}
\end{figure}

A total of 123 {\ha} emitters were selected among the 1242 SAMI sources within the MUSE FOV, corresponding to $10 \pm 1\%$\footnote{The fractional errors shown throughout the manuscript were derived using a Poisson distribution, i.e., $\sigma_\mathrm{Be}=N_\mathrm{Be}^{1/2}/N_\mathrm{total}$.} of the sample. This fraction is much smaller than the $32 \pm 3\%$ reported by \citet{Bodensteiner20} based on a flux-limited sample of stars.
The bright subsample ($V<18.5$~mag) of SAMI sources provides a direct comparison with their results. Here, a total of 87 {\ha} emitters were identified. A fraction of $27 \pm 3$\% {\ha} emitters was observed after the exclusion of 16 giant stars located on the right side of the diagram ($V-I > 0.5$~mag, middle panel of Fig.~\ref{fig_ccd_ngc330_Ha_sel_museFOV}). 
Finally, 63 CBe stars + 10 non-CBe stars from the MUSE sample exhibited {\ha} excess from 250 targets. By excluding the giants located on the right side of the diagram (14), we end up with a fraction of $31 \pm 4\%$ of {\ha emitters} for the MUSE samples, which is in good agreement with the findings from \citet[][$32\pm3$\%]{Bodensteiner20}.

The analysis of the flux-limited SAMI subsample led to a similar fraction of CBe stars ($27\pm3$\%) as that obtained from spectroscopic MUSE observations ($32\pm3$\%). Despite that, the deeper SAMI photometry allowed us to identify a significant number of fainter {\ha} emitters, consistent with late-type CBe stars candidates. As the population of normal stars increases exponentially as a function of the mass (and the $V$-band magnitude), the total fraction of active CBe stars with clear {\ha} emission decreases toward late-B stars, reaching $10\pm1$\% when considering all stars exhibiting $V \leq 22$~mag.

\subsection{Fraction of \texorpdfstring{\ha}{H-alpha} emitters versus normal stars}

The C-M diagrams of Fig.~\ref{fig_cmd_ngc330} show that {\ha} emitters are found over a wide range of luminosities ($12 < V \lesssim 22.0$~mag). This range is wider than that expected for B-type stars ($13.6 \leq V \leq 20.2$~mag), indicating the SAMI photometry is deep enough to identify all CBe stars exhibiting strong enough {\ha} emission within NGC\,330.

By taking advantage of the deeper photometric completeness of SAMI photometry, we can now obtain a complete census of the fraction of \ha emitters stars versus normal MS stars associated with NGC\,330.
For the analysis presented in this section, we selected a total of 3318 sources exhibiting $B-I < 0.5$~mag, i.e., excluding the bright evolved objects and the RGB sources (see Fig.~\ref{fig_cmd_ccd_ngc330_isochrone}).

The next subsections present the analysis of the fraction of {\ha} emitters as a function of
$i)$ the distance from the center of NGC\,330, and 
$ii)$ the spectral type of B stars.

\subsubsection{Radial distribution of \texorpdfstring{\ha}{H-alpha} emitters in \texorpdfstring{NGC\,330}{NGC330}}

We investigated the surface density of {\ha} emitters as a function of the distance from the cluster's center (assuming as central coordinates $\alpha = 00$h\,$56$m\,$16.9$s, $\delta = -72^{\circ}\,27\arcmin\,48\arcsec$).
For reference and comparison, we also computed the surface density of MS objects (sources exhibiting $B-I < 0.5$~mag and not selected using Eq.~(\ref{eq_Ha_selection}), dividing them into higher- and lower-mass MS stars with respect to $V \leq 18.0$~mag (corresponding to the transition between B5 and B6 stars) to look for evidence of mass stratification within the cluster.
The distance of 57.5~kpc to the SMC was adopted \citep{milone2018} to convert the angular projected distances from arcseconds to parsecs.

The resulting curves are presented in the top panel of Fig.~\ref{fig_be_fraction_radius}, showing the number of sources per parsec${^2}$, evaluated at concentric rings from 0 to {140\arcsec}, in steps of {5\arcsec} (solid lines). 
For comparison, the cumulative distribution function (CDF) of the total number of stars as a function of the projected distance is shown as the dashed lines in the same plot.

\begin{figure}[!ht]
\centering
\includegraphics[width=\linewidth]{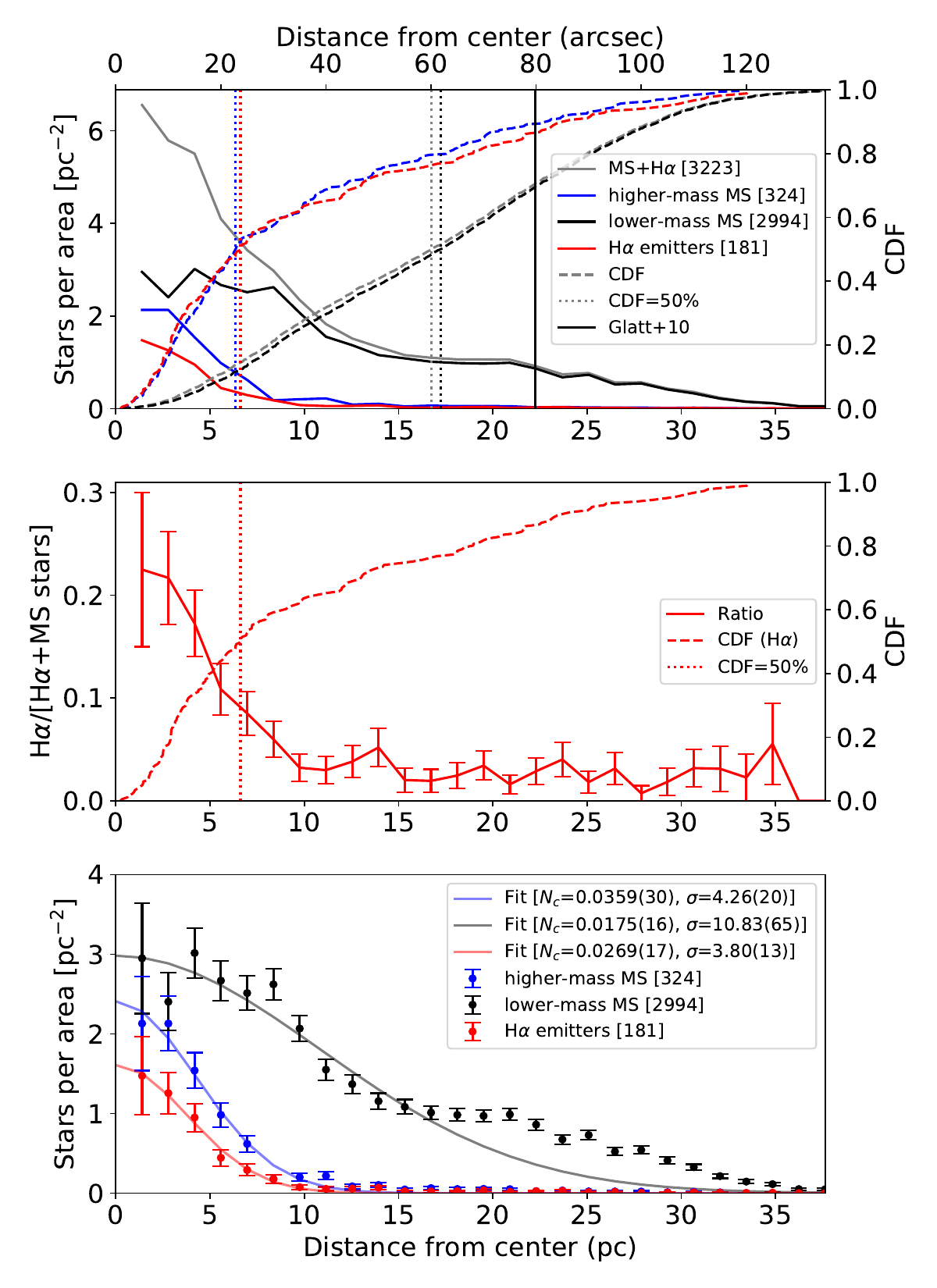}\\[-2.0ex]
\caption{Top panel: Surface density (number of sources per area) of normal MS stars (the filled blue curve indicates higher-mass objects associated with $V\leq18$ mag, and fainter, lower-mass objects by the filled black curve), {\ha} emitters (filled red), and the total number MS and \ha emitters (filled grey) as a function of the distance from the center of NGC\,330.
The values were calculated for concentric rings around the central coordinates of the cluster in steps of {5\arcsec}.
The dashed curves indicate the cumulative distribution function of the quantities, while the dotted curves indicate the distance corresponding to $\textrm{CDF} = 0.5$. For reference, the vertical filled black line indicates the 1\farcs33 cluster radius from \citet{Glatt10}.
Middle: The {\ha} emitter fraction as a function of the distance from the center of NGC\,330 (solid line). {The uncertainties are indicated by the error bars.} For comparison, the CDF for {\ha} emitters is indicated by the dashed curve.
Bottom: Three-dimensional Gaussian fit model to the surface density of {\ha} emitters (red), {higher- (blue) and lower-mass MS stars (black)}. The best parameters for each curve ($N_c$, $\sigma$) are indicated on the label.
\label{fig_be_fraction_radius}}
\end{figure}

The surface density of the sources with available SAMI photometry (filled grey curve) shows a cusp pattern towards the center of the cluster, decreasing towards larger radii. The CDF (dashed grey curve) reaches 50\% at a radius of 16.8~pc ($\sim$60\arcsec), which is about 75\% smaller than the radius of the cluster reported by \citet[][indicated by the horizontal filled black line]{Glatt10}.
The lower-mass MS stars exhibit a similar surface density profile (black curve), and the CDF of the total number of low-mass stars reaches 50\% at a projected distance of 17.2~pc (62\arcsec), roughly following the CDF of the MS+{\ha} sample (in grey).
The higher-mass MS stars (blue curve), however, exhibit a much more compact distribution, indicating that massive stars are preferably located within the inner region of the cluster (the CDF reaches 50\% at 6.4~pc (23\arcsec).
Finally, the {\ha} emitters (red curves) exhibit a similar distribution as that observed for the high-mass MS stars, and its CDF reaches 50\% at 6.6~pc (24\arcsec) distance. These results suggest that the {\ha} emitters and the massive stellar populations are likely confined within the inner region of the cluster.

The middle panel of Fig.~\ref{fig_be_fraction_radius} shows the fraction of {\ha} emitters, i.e., {\ha}/(MS+\ha), as a function of the distance to the cluster's center.
The data suggests a significant increase in the proportion of \ha emitters towards the center, in agreement with the surface density results above. About 50\% of all {\ha} emitters and $>50\%$ of the higher-mass MS stars are concentrated within a circle of 6.6~pc (24\arcsec) radius, whereas only $\sim10\%$ of the lower-mass MS objects are found within the same area. Such distinct distribution between higher- and lower-mass objects corresponds to a clear evidence of mass segregation within NGC\,330.

The stellar surface density measurements can be used to infer the physical properties of the cluster{. For example, they provide } valuable information about the structure, density profile, and dynamical characteristics, offering insights into the formation and evolution of stellar systems and populations.
Taking the analysis a step further, we transformed the surface density information mentioned earlier into volume density. This conversion assumes that the stars exhibit a spatial distribution following a three-dimensional Gaussian distribution, as detailed in Appendix~\ref{appendix_gauss3d}).
For this analysis, we can estimate the stellar density at the center of the cluster (in stars~pc$^{-3}$) and the characteristic scale (i.e., standard deviation, $\sigma$) of the stellar population (in pc). 

The bottom panel of Fig.~\ref{fig_be_fraction_radius} presents the observed stellar density of the MS stars and {\ha} emitters as a function of the projected distance from the center of the cluster, together with the best three-dimensional Gaussian fit. We observe a reasonably good fit for all datasets for $r \lesssim 15$~pc, indicating that the model can reproduce the observed spatial stellar distribution. This simple model better fits the observed dataset with greater accuracy when compared to, e.g., simplified King profiles often adopted for young stellar clusters \citep[e.g.][]{Bonatto09,Hetem19}. Therefore, adopting this profile is encouraging and warrants testing with other young stellar clusters.
The characteristic scale for the lower-mass MS stars is $\sigma=10.83\pm0.65$~pc, while the $\sigma$ values are relatively shorter for the higher-mass and {\ha} emitters ($4.26\pm0.20$ and $3.80\pm0.13$~pc, respectively). These results show unequivocally that the {\ha} emitters and the most massive stars are preferably located in a more compact region from the center of the cluster. At the same time, low-mass MS objects are observed towards larger distances.

In addition, the central density for high-mass MS and {\ha} emitters corresponds to $N_c = (35.9 \pm 0.30) \cdot 10^{-2}$~pc$^{-3}$ and $(26.9 \pm 0.17) \cdot 10^{-2}$~pc$^{-3}$, respectively, while the density of lower-mass MS stars is relatively lower, $N_c = (17.5 \pm 0.16) \cdot 10^{-2}$~pc$^{-3}$. These estimates quantitatively support
$i)$ the existence of a significant mass stratification within NGC\,330, and
$ii)$ that {\ha} emitters are preferably found towards the inner region of the cluster. 

\subsubsection{Fraction of \texorpdfstring{\ha}{H-alpha} emitters as a function of the Spectral Type}

We further investigated the fraction of {\ha} emitters as a function of the $V$-band magnitude (i.e., spectral type) to assess its distribution vs. stellar mass within NGC\,330 and in the adjacent stellar field. We defined two sub-samples of normal MS stars and {\ha} emitters exhibiting with $B-I<0.5$~mag (therefore excluding older objects). This selection led to 3397  highly probable members located within the radius of NGC\,330, $r={1.33\arcmin}$ \citep{Glatt10}.
In an effort to examine the B-type stars associated with {\ha} emission, we selected a subset of 858 sources exhibiting $V$-band magnitudes within the range expected for B0-B9 stars (from Table~\ref{table_magnitudes_smc}. 
Additionally, 1631 sources were identified at greater distances from the cluster's center, which are more likely to be field stars.
We adopted the Main Sequence Turn Off point (MSTO) as $V = 15.92$~mag (in AB magnitudes) from \citet{Iqbal13}, equivalent to the age of NGC\,330 estimated as $\log(\textrm{age})= 7.4$ dex ($\sim25$~Myr) from \citet{Glatt10}.
The results are presented in Fig.~\ref{fig_be_fraction_v_sep}.

\begin{figure*}[!ht]
\centering
\includegraphics[width=\linewidth]{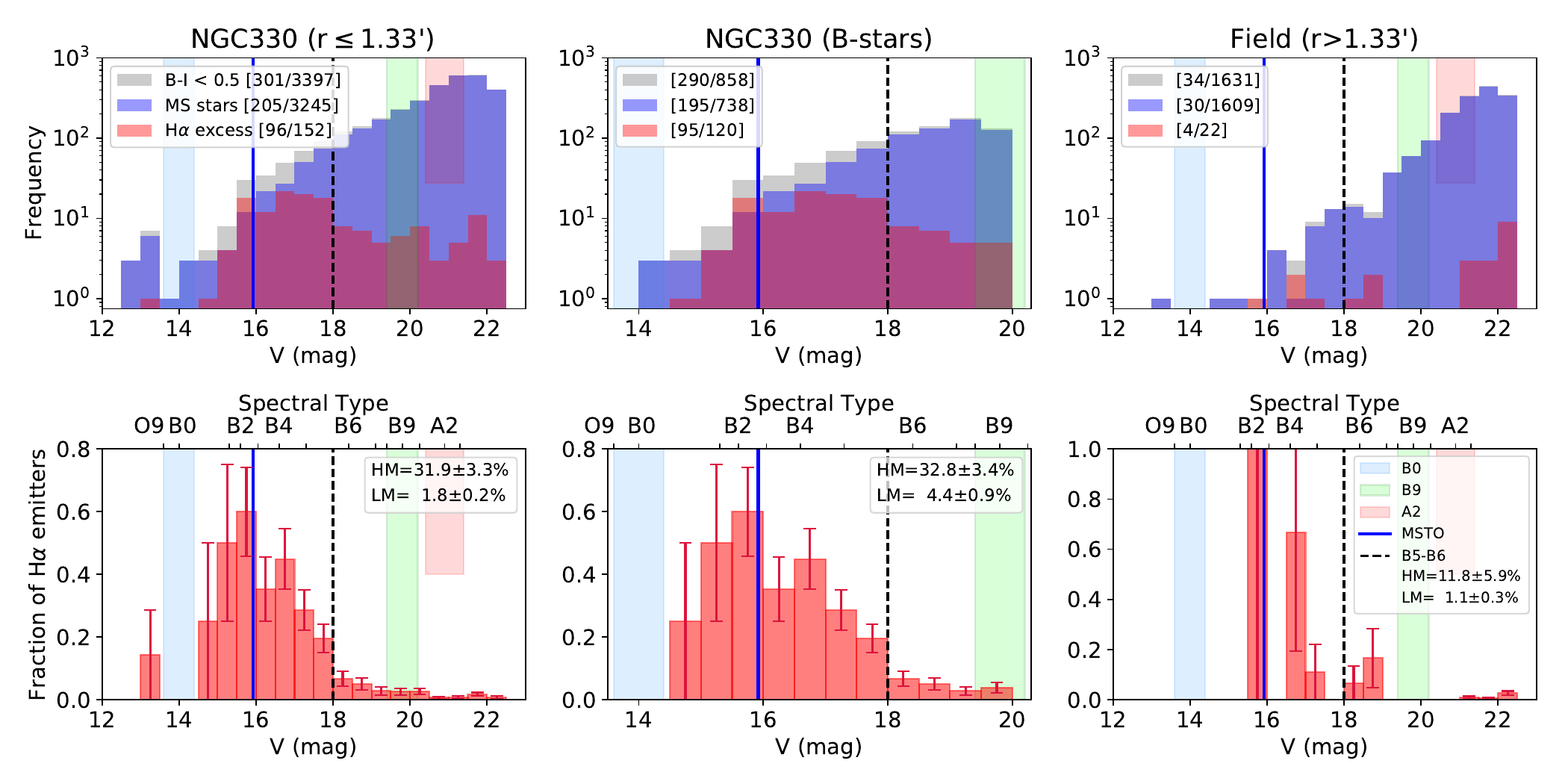}\\[-3.0ex]
\caption{Top panels: Distribution of normal MS stars and {\ha} emitters as a function of the $V$-band magnitude. From left to right, we show the total number of stars within a 1\farcs33 radius from the center of NGC\,330, the B-type star candidates in NGC\,330, and the total number of stars in the field ($r > 1\farcs33$).
The expected magnitude intervals for B0 (blue), B9 (green), and A2 stars (red) are indicated by the shaded regions.
The vertical filled blue line indicates the MSTO ($V=15.92$) for NGC\,330, and the vertical dashed black lines divide the sample into higher- (HM) and lower-mass stars (LM) using $V=18.0$~mag as reference.
The values on the legend correspond to the number of sources brighter than $V = 18.0$, and the total number of sources, respectively.
Bottom: the ratio of \ha emitters a function of the $V$-band magnitude for the same subsamples presented in the top panel.
The upper axis indicates the mean $V$-band magnitude for each spectral type listed in Table~\ref{table_magnitudes_smc}. The ratio of {\ha} emitters for the HM and LM subsamples are indicated on each panel.
The error bars correspond to the fractional errors for each bin.
\label{fig_be_fraction_v_sep}}
\end{figure*}

The top left panel of Fig.~\ref{fig_be_fraction_v_sep} presents the distribution of the normal MS stars (blue bars) and {\ha} emitters (red) as a function of their $V$-band magnitude.
For comparison, the associated spectral sub-types are also indicated in the plots, using the mean value of the magnitudes listed in Table~\ref{table_magnitudes_smc}.
The sources are distributed across a wide range of brightness ($12 \leq  V \leq 22.5$), covering the entire magnitude interval expected for B-type stars.
While the number of normal MS objects increases towards late-type objects following a power law (i.e., IMF-like), the distribution of {\ha} emitters peaks at $V$ between 15.5 and 18~mag (B1-B5 stars), and the bulk of {\ha} emitters is slight offset towards fainter magnitudes than the MSTO for NGC\,330.
A total of 152 {\ha} emitters are observed within the entire $V$-band magnitude range, corresponding to a fraction of 4.4$\pm$0.5\% of {\ha} emitters ({\ha}/[\ha+MS]) with $V \lesssim 22.5$~mag.

The bottom left panel of Fig.~\ref{fig_be_fraction_v_sep} shows the fraction of {\ha} emitters as a function of the $V$-band magnitude. As observed on the top panel, the largest fraction of {\ha} emitters is also found for $V$-band magnitudes between 15 and 18~mag, consistent with B1-B5 stars.
The distribution reaches a peak of $60\pm14\%$ at $V \sim 15.5$~mag, corresponding to early B-type objects (B1-B3) close to the MSTO, in agreement with findings from previous seeing-limited results from \citet{Iqbal13}.

The fraction of {\ha} emitters decreases to values smaller than 10\% for $V > 18$~mag (i.e., for spectral sub-types later or equal to B6).
Despite the smaller fraction observed towards fainter objects, the total number of sources with {\ha} excess (top panel) is remarkable, indicating a considerable number of late-B type stars exhibiting {\ha} emission, a result absent in previous works due to the relatively faint nature of such objects.
To properly compare our statistics with previous works, we adopted $V=18.0$~mag (corresponding to B5-B6 stars) as a limit to estimate the fraction of {\ha} emitters for higher- and lower-mass stars (indicated as HM and LM in the legend of the figure). Such division led to a fraction of $31.9\pm3.3\%$ of {\ha} emitters for stars more massive than B5-B6, and $1.8\pm0.2\%$ for lower mass objects.

A similar behavior is observed when restricting the analysis towards the expected magnitudes for B-type stars ($13.6 \leq V \leq 20.2$, from Table~\ref{table_magnitudes_smc}), as indicated by the middle panels in Fig.~\ref{fig_be_fraction_v_sep}. For such sub-sample, the {\ha} emitter fraction reaches a peak of $60\pm15\%$ for B2 stars at the limit of the MSTO, decreasing towards late-type objects. The fractions are slightly larger than observed for the full sample (left panel), reaching $32.8\pm3.4\%$ and $4.4\pm0.9\%$ for the higher- and lower-mass samples, respectively.
The higher fraction of {\ha} emitters towards the more massive stellar population agrees with the previous results from \citet{Bodensteiner20}, while the lower fraction towards lower-mass objects is an interesting new result, indicating a small but significant fraction of active stars towards the lower mass limit of B-type stars.
{The decrease of the CBe fraction with decreasing mass has been pointed out by several authors \citep[e.g.,][]{Keller99, Iqbal13, Bodensteiner20, Hastings21}; however, this is the first time a non-zero fraction of {\ha} emitters is actually observed at the lower-mass limit of the B-type population in the SMC.
In particular, \citet{Hastings21} observed a sudden decrease in the CBe fraction beyond spectral types later than B5 in NGC\,330 (e.g., see their Fig.~3). We argue that may be influenced by the flux limitations of their {\ha} observations (based on HST photometry from \citealt{milone2018}).} 

For completeness, the top right panel of Fig.~\ref{fig_be_fraction_v_sep} presents the distribution of {\ha} emitters and MS stars for the field stars. The MS stars exhibit a similar IMF-like distribution as observed 
on the top left panel, with very few bright (e.g., massive) objects exhibiting $V < 18$~mag. This is expected as massive stars (at the top of the IMF) are less likely to be isolated stars, i.e., unbound to the cluster.
Only 22 of the 1631 sources at larger radii from NGC\,330 are {\ha} emitters. While most of them are fainter than $V=21$~mag, only four are brighter than $V=18$~mag.
Consequently, low-count statistics at the high-mass end likely affect the fraction of {\ha} emitters, as indicated by the bottom right panel of Fig.~\ref{fig_be_fraction_v_sep}.
The integrated high-mass fraction ($11.8\pm5.9\%$) is about three times lower than observed towards NGC\,330, while the fraction remains at $\sim$1\% values for the lower-mass objects.

\section{Discussion and Conclusions}
\label{sec_discussion}

{Although previous studies have investigated the B content of NGC\,330 with high-angular resolution observations (e.g., \citealt{milone2018}, \citealt{Bodensteiner20}), this work represents the first study combining high-angular resolution with sufficient photometric depth to investigate the stellar population up to early A-type ($V \approx 22$~mag) in both broad- $(\mli{BVRI})$ and narrow-band {(\ha)} filters.}
The unprecedented quality of AO-assisted SAMI/SOAR observations allowed us to resolve the stellar population of NGC\,330 at scales of $\gtrsim0\farcs1$ in all five filters, and to identify {\ha} emitters throughout the entire B-type stellar population, providing an excellent opportunity to investigate the Be phenomenon and the occurrence of CBe stars across all spectral sub-classes.

We first established a photometric calibration based on a \emph{stellar template model fitting} methodology to calibrate all $\mli{BVRI}$+H$\alpha$ filters of the SAMI catalog, delivering photometric errors smaller than 0.3~mag for sources at the 23-mag limit using less than a hundred known objects. As a direct consequence of using stellar template models, we also obtained the corresponding stellar parameters for each object, and these findings will be utilized in an upcoming publication.

The astrometric information from the Gaia DR3 catalog was used to identify and filter out 849 foreground objects in the line-of-sight of NGC\,330. In addition, using a simple inspection of PARSEC evolutionary tracks, we identified 674 old Red Giant Branch stars that are unlikely to be associated with the young stellar cluster, removing them from further analysis. {The online version of Table~\ref{table_phot_SAMI} presents the classification of the full SAMI catalog.}


Through the adoption of a color selection criterion based on color indices, we successfully identified 181 {\ha} emitters within the NGC\,330 field. Among these sources, 137 are brighter than $V<20.2$~mag, corresponding to stars equal to or earlier than B9. This count surpasses by a factor of two the previously reported number of {\ha} emitters in literature, based on similar selection criteria \citep{Keller99,Iqbal13}. The remaining 44 {\ha} emitters correspond to fainter, lower-mass objects within the magnitude range of $20.2 < V \lesssim 22.5$~mag.

{We conducted a comparison between the AO-assisted SAMI observations and the recent HST observations reported by \citet{milone2018}. They identified 63 {\ha} emitters consistent with main-sequence objects in NGC\,330. Even with their analysis relying on high-angular resolution observations, they are constrained to {\ha} emitters exhibiting $I$~$\sim$~18~mag, roughly equivalent to B6 objects. We conducted a direct comparison using a flux-limited sample of objects exhibiting $I$~$<$~18~mag on the SAMI/SOAR photometric catalog (Table 3), resulting in a total of 83 {\ha} emitters. This population is 30\% larger than that reported by \citet{milone2018}.}

{Similarly, we compared our SAMI observations with those reported by \citet{Bodensteiner20} based on MUSE observations.}
Our focus was on the identical sky region, and we restricted our sample to match their magnitude limit ($V<18.5$~mag). Based on spectroscopic classification, \citeauthor{Bodensteiner20} identified a total of 82 stars (2 Oe and 80 Be) leading to a Be fraction of Be/(B+Be)~=~$32\pm3\%$.
Using our adopted color selection criterion (Eq.~\ref{eq_Ha_selection}), we identified 87 stars in our magnitude-limited sample, leading to a fraction of {\ha} emitters of $27\pm3\%$, in good agreement with the previous study.

Further comparison was made with the previous seeing-limited works of \citet{Keller99} and \citet{Iqbal13}, who analyzed the fraction of CBe stars for a flux-limited sample of objects exhibiting $V<17$~mag.
\citet{Keller99} reported that the fraction of CBe stars reaches a maximum of 80\% at $V$ between 15.0 and 15.5~mag (consistent with B1-B2 objects), decreasing towards fainter objects. Their findings are very similar to the results presented in Fig.~\ref{fig_be_fraction_v_sep}.
In addition, \citet{Iqbal13} found the fraction of \ha emitters peaks at the MSTO ($V \sim 16$~mag), at 40\% (see their Fig.~4). Although their fraction is smaller than observed from the SAMI observations (see below), our observations also match the peak of CBe stars closer to the MSTO (cf. their Fig.~4).
The existence of stars above the MSTO can only be explained by binary interaction or rapid rotation, therefore explaining the significantly large fraction of CBe stars for stars earlier than B2-B3.
{This finding is consistent with the results of \citet{georgy2013}, who estimated an increase in the MS lifetime by factors of up to 35\% depending on the  initial rotation and mass of a star (at solar metallicity), with a potential extension of up to 60\% for SMC metallicities (Z=0.002).}

Our comprehensive sample enables us to construct a detailed perspective of the \ha-emitting population in NGC\,330. We demonstrate that {\ha} emitters are found across a broader range of magnitudes than that expected for B stars, signifying that \ha emitters are found across the late O up to A spectral subtypes.

An analysis of the radial spread of the stars in the cluster revealed that the surface density distribution of normal MS stars is clearly bimodal: the higher-mass stars (arbitrarily defined as the ones brighter than $V=18$~mag, roughly earlier than B6) are concentrated in the inner part of the cluster. In contrast, the lower-mass ones are spread over a much larger volume. Furthermore, the \ha emitting stars are spatially correlated with the higher-mass ones. From the above, and further supported by Fig.~\ref{fig_be_fraction_v_sep}, two important conclusions may be drawn:
$1)$ the \ha emitters are much more frequent among the higher-mass subsample, reaching a peak value of almost 60\% for B2 stars;
$2)$ the stellar content of the cluster clearly displays mass stratification, with the higher-mass ones (presumably also with a high binary content, e.g., \citealt{Sana14}) occupying a much smaller effective volume. 
The current analysis offers a more robust confirmation of mass stratification within NGC\,330 compared to earlier space-based HST observations indicating similar evidence (e.g., \citealt{Sirianni02, Gouliermis04}).

When the whole main-sequence population is considered (removing the stars with $B-I> 0.5$~mag), we found a {\ha}/({\ha}+MS) ratio of $4.4\pm0.5\%$. This is much lower than the fraction of CBe stars reported by \citet[][$34\pm8\%$]{Keller99} and \citet[][16.6\%]{Iqbal13}. The reason is apparent from Fig.~\ref{fig_be_fraction_v_sep}, which shows that the total number of stars rises quickly towards later spectral types while the number of {\ha} emitters drops sharply. When we restrict the analysis to the B-stars (middle panels of Fig.~\ref{fig_be_fraction_v_sep}), the fraction increases to $14.0\pm1.3\%$, but is still much lower than the magnitude-limited estimates from the previous works.
The fraction of {\ha} emitters significantly increases when restricting the analysis towards the stars brighter than $V=18$~mag (corresponding to B5-B6 stars), showing {\ha} fractions around 32\%, which is consistent with the fraction of CBe stars reported by \citet[][$32\pm3\%$]{Bodensteiner20}.

The photometric analysis presented in this work is biased towards {\ha} emitters with relatively strong H$\alpha-R$ excess (Fig.~\ref{fig_ccd_ngc330_Ha_sel}).
Such \ha excess may unequivocally be associated with CBes that were active at the time of the observations and, thus, possessed large and dense disks.
Conversely, we expect that weak {\ha} emitters or quiescent CBe stars with proper spectroscopic classification from \citet{Bodensteiner20} are generally not selected by the color criterion. Indeed, ten of the CBe stars classified by those authors do not satisfy Eq.~\ref{eq_Ha_selection} (indicated as red $\times$ symbols in Fig.~\ref{fig_ccd_ngc330_Ha_sel_museFOV}). In contrast, ten of their B-type stars are clear {\ha} emitters in our sample (indicated as open cyan circles in the same figure). 

Despite the large dispersion on the fraction of {\ha} emitters as a function of mass, a significant number of individual {\ha} emitters are observed over the entire range of magnitudes shown in Fig.~\ref{fig_be_fraction_v_sep}. On the premise that {\ha} emission is a {strong} indicator of CBe activity, this evidence shows that CBe stars {exist} regardless of their main sequence lifetime fraction ($t/t_{\mathrm{MS}}$), supporting that the Be phenomenon observed in NGC\,330 is unlikely to be mainly ruled by evolutionary phases (such as core contraction), in agreement with the findings from \citet{Iqbal13,Keller99}.

Even though the telltale signature of the Be phenomenon is {\ha} emission, estimating the $\mathrm{Be}/(\mathrm{B}+\mathrm{Be})$ fraction using only photometry is a challenging task, as there are many foreseeable situations in which CBe stars will not display a detectable \ha$-R$ color excess. For instance, during the early (late) stages of disk construction (dissipation) the line emission can be too weak. Also, some CBe stars only develop tenuous disks at times \citep[e.g.,][]{2017MNRAS.464.3071V}. Finally, many CBes are seen without any traces of disks for extended periods of time. One classical example is $\delta$~Sco, which was considered a normal B-type star for over a century, but grew a disk in the year 2000 that lasts to this day \citep{2020ApJ...890...86S}.

From the above, it follows that the real $\mathrm{Be}/(\mathrm{B}+\mathrm{Be})$ fraction will always be higher than the observed {\ha} emitter fraction in young clusters.
One way to improve estimates of the CBe content is to consider observations at different epochs.
For instance, when considering the MUSE objects with available SAMI photometry from Fig.~\ref{fig_ccd_ngc330_Ha_sel_museFOV}, ten spectroscopically classified CBe stars do not show {\ha} excess. In contrast, 63 CBe stars and ten B stars are identified as {\ha} emitters. When considering all these 83 objects, the lower limit of the $\mathrm{Be}/(\mathrm{B}+\mathrm{Be})$ fraction would slightly rise from $31\pm4\%$ to $35\pm4\%$.

In the following paper of this series, a new methodology will be presented, taking into account the complexities of CBe stars to create realistic stellar populations and their ensuing photometry to obtain a much better estimate of the CBe content of NGC\,330, without the {known} biases outlined above.

\begin{acknowledgments}
{We thank the referee for the useful comments and suggestions that led to a much improved version of this manuscript.}
The work of F.\,N.\, is supported by NOIRLab, which is managed by the Association of Universities for Research in Astronomy (AURA) under a cooperative agreement with the National Science Foundation.
P.\,T.\, acknowledges support from CAPES (grant 88887.604774/2021-00).
A.\,C.\,C. acknowledges support from CNPq (grant 311446/2019-1) and FAPESP (grants 2018/04055-8 and 2019/13354-1). A.\,F.\, acknowledges support from CAPES (grant 88882.332925/2019-01) and FAPESP (grant 	
2023/06539-0).
This work made use of the computing facilities of the Laboratory of Astroinformatics (IAG/USP, NAT/Unicsul), whose purchase was made possible by the Brazilian agency FAPESP (grant 2009/54006-4) and the \mbox{INCT-A}.
Based on observations obtained at the Southern Astrophysical Research (SOAR) telescope, which is a joint project of the Minist\'{e}rio da Ci\^{e}ncia, Tecnologia e Inova\c{c}\~{o}es (MCTI/LNA) do Brasil, the US National Science Foundation’s NOIRLab, the University of North Carolina at Chapel Hill (UNC), and Michigan State University (MSU).

\end{acknowledgments}


\vspace{5mm}
\facilities{SOAR(SAMI)}


\software{Astropy \citep{2013A&A...558A..33A,2018AJ....156..123A},
          SciPy \citep{Virtanen20},
          IDL
          }


\bibliographystyle{aasjournal}


\appendix

\section{Comparison between SAMI photometry and external catalogs}
\label{appendix_catalogs}

For completeness of the results exhibited in Sect.\,\ref{sect_photometric_comparison}, the comparison between the SAMI photometry and the external catalogs from MCPS and MUSE is presented as follows.

\subsection{MCPS}
\label{appendix_mcps}

We compared the calibrated $\mli{BVI}$ magnitudes with those from the MCPS catalog for sources brighter than 20~mag for all filters.
The residuals between SAMI and MCPS magnitudes are presented in the left panel of Fig.~\ref{fig_photcal_residuals_MCPS_MUSE}.

The residuals peak around 0~mag, exhibiting an extended tail towards positive values. Therefore, the 1-$\sigma$ widths of the distributions are relatively larger than those observed for OGLE (see Fig.~\ref{fig_photcal_residuals_OGLE}).
According to Fig.~\ref{fig_minsep}, the minimum distance of the nearest neighbor for MCPS is 1\farcs04, the largest value observed among all the other photometric catalogs, suggesting that a single and bright object in MCPS is likely resolved into multiple and fainter objects in SAMI observations, explaining the positive residuals as observed in Fig.~\ref{fig_photcal_residuals_MCPS_MUSE}.

\begin{figure}[!ht]
\centering
\includegraphics[width=0.475\linewidth]{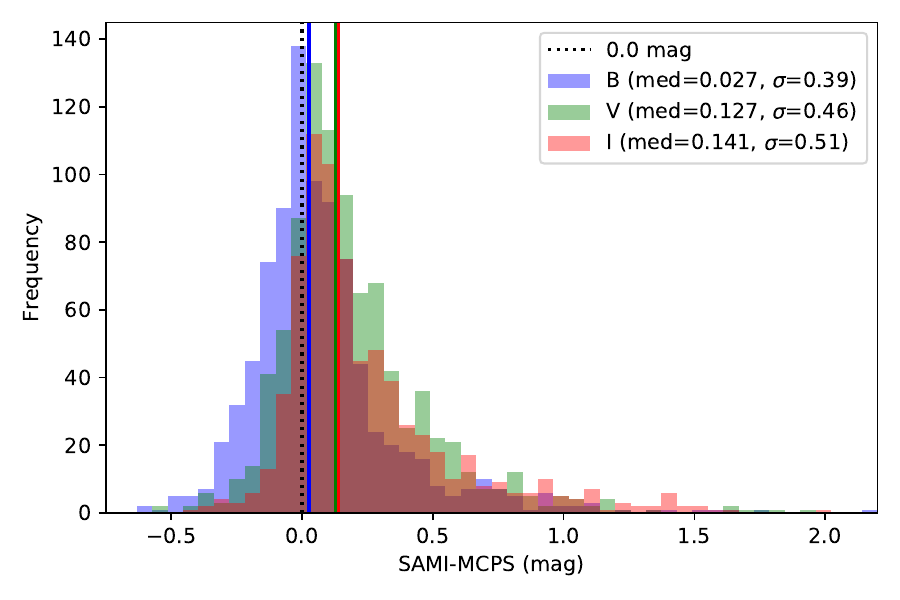}
\includegraphics[width=0.475\linewidth]{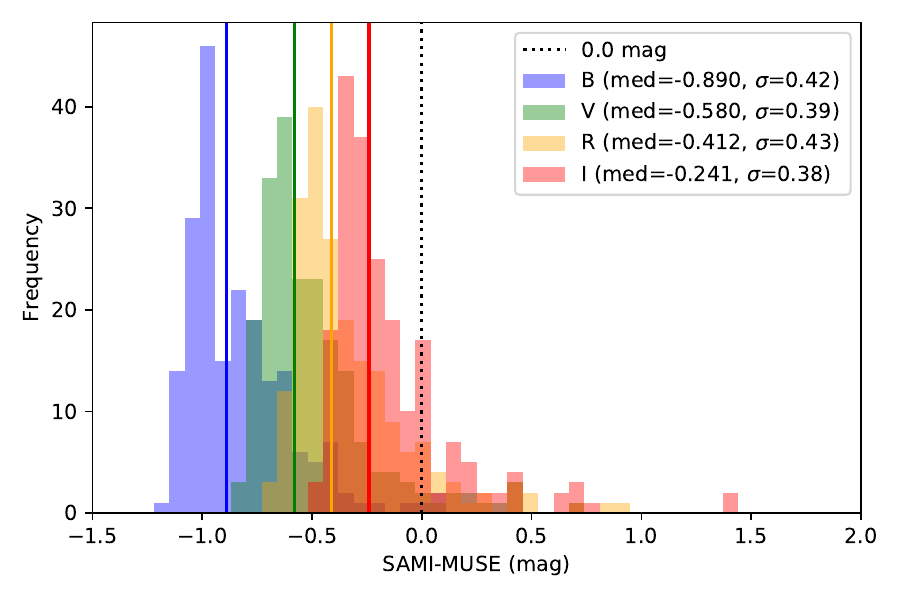}\\[-3.0ex]
\caption{Left panel: Residuals between SAMI and MCPS photometry. Right panel: Residuals between SAMI and MUSE photometry. The full description of the plots is given in Fig.\,\ref{fig_photcal_residuals_OGLE}.
\label{fig_photcal_residuals_MCPS_MUSE}}
\end{figure}

\subsection{MUSE}
\label{appendix_muse}

The right panel of Fig.~\ref{fig_photcal_residuals_MCPS_MUSE} presents the residuals between the $\mli{BVRI}$ magnitudes from SAMI and MUSE photometry. The synthetic MUSE photometry was obtained by extracting the one-dimensional spectrum of the objects and convolving them with the SAMI filter transmissions. The distribution of the photometric residuals exhibits a systematic negative offset towards the bluest filters, ranging from $-0.241$~mag for the $I$-band to $-0.89$~mag to the $B$-band. Given that the comparison between OGLE (Sect.~\ref{sect_photometric_comparison}) and MCPS (Appendix~\ref{appendix_mcps}) do not exhibit any systematic divergence, we speculate that a zero-point correction on the flux-calibrated MUSE datacube can be the cause of such large offsets.

\section{A three-dimensional Gaussian Model for the Stellar Density}
\label{appendix_gauss3d}

The formula for a 3D Gaussian function in cylindrical coordinates is given by:

\begin{equation}
n(r, \phi, z) = N_{\rm c} \cdot \exp\left(-\frac{r^2}{2 \sigma_r^2} - \frac{z^2}{2 \sigma_z^2}\right)\,,
\end{equation}
where it was assumed, for simplicity, symmetry around the $z$-axis. $N_c$ is the density of stars (stars\,$\rm pc^{-3}$) at the cluster's center.
Further assuming that the observer is placed along the $z$-axis, and that the standard deviations along the $r$ and $z$ directions are the same (i.e., $\sigma_r=\sigma_z=\sigma$), we have that the observed superficial density  of stars, $\Sigma(r)$ (in pc$^{-2}$), at a distance $r$ from the cluster's center is
\begin{equation}
    \Sigma(r) = \int_{-\infty}^{\infty} n(r,\phi,z) dz
    \int_{0}^{2\pi} d\phi
    =
    (2\pi)^{3/2} N_{\rm c} \cdot \sigma \cdot
    \exp\left(-\frac{r^2}{2 \sigma^2}\right)\,.
\end{equation}



\end{document}